\documentclass[12pt, letterpaper]{extarticle}

\usepackage{subcaption}
\usepackage{fullpage}
\usepackage[switch]{lineno}
\usepackage{amsmath}
\usepackage{amssymb}
\usepackage{rotating}
\usepackage{array}
\usepackage{mathtools}
\usepackage[ruled]{algorithm2e}
\usepackage{algorithmic}
\usepackage{bm}
\usepackage{breqn}
\usepackage{comment}
\usepackage{enumitem}
\usepackage{graphics}
\usepackage{graphicx}
\usepackage{latexsym}
\usepackage{mathrsfs}
\usepackage{morefloats}
\usepackage{nicefrac}
\usepackage{authblk}
\usepackage{pifont}
\usepackage{times}
\usepackage{xcolor}
\usepackage{placeins}
\usepackage[numbers,sort&compress]{natbib}

\usepackage[hyphens]{url}
\usepackage{hyperref}
\hypersetup{colorlinks=false,breaklinks=true}

\usepackage{etoolbox}
\AtBeginEnvironment{quote}{\par\singlespacing\small}


\title{What Societies Are Polarized About: Political Attitudes on Five Dimensions in 155 Countries}

\author[1]{Hazem Ibrahim}
\author[1]{Aymane Omari}
\author[2*]{Aaron R. Kaufman}

\affil[1]{\normalsize Computer Science, Science Division, New York University Abu Dhabi, UAE.}
\affil[2]{\normalsize Political Science, Social Science Division, New York University Abu Dhabi, UAE.}
\affil[*]{\footnotesize Corresponding author. E-mail: aaronkaufman@nyu.edu}
\date{}

\begin{document}

\maketitle

\baselineskip22pt

\begin{abstract}
Societies worldwide are described as increasingly polarized, yet issue polarization across countries has been measured almost entirely on a single left--right axis developed in Western democracies. Consequently, countries divided on another axis will appear unpolarized. Here, we harmonize 1,378 items from twelve cross-national survey programs and national election studies in 17 countries to place 2.13 million people interviewed in 155 countries since 2013 on five dimensions: social liberalism, economic left--right, democratic commitment, anti-elite sentiment, and the role of religion in law and government. National publics differ sharply in what divides them. European countries divide most often over social liberalism, Middle Eastern and North African countries over democratic commitment, Sub-Saharan African countries over anti-elite sentiment, and Latin American countries over economic left--right. Religion in government, the cleavage most associated with the Middle East's Islamist--secular divide, is the sharpest division in ten countries, four of them in Europe and Central Asia, four in East Asia and the Pacific, and none in the Middle East. The dimensions also correlate poorly within people: the median correlation between a person's social and economic positions is $0.01$, and only 24 of the 96 countries we test exceed $+0.1$, led by the United States at $+0.54$. The dimension that divides a public most also divides its parties most in fewer than half of countries, though twice as often as chance; and despite research attention to gender, age, and urban--rural divides, the demographic gaps on these dimensions are small and heterogeneous across different countries. We conclude that understanding political conflict requires asking not how polarized a society is, but what it is polarized about.
\end{abstract}

\clearpage

\section*{Introduction}

Societies around the world are described as increasingly polarized \citep{falkenberg2026toward}, yet the instruments used to evaluate that claim in mass publics across countries are almost always single left--right scales developed in and for Western democracies. Research on party competition in Europe has long worked with two dimensions, one economic and one cultural, and finds that how they combine differs from country to country \citep{kriesi2006globalization,rovny2019new}; cross-national measures of mass polarization have not followed. An instrument built around one region's political divisions determines, by construction, which divisions elsewhere are visible at all, and a country may appear undivided simply because it is divided about something its questionnaire does not address. Iraq, for instance, sits at the 15th percentile of the world's countries in how divided it is on social issues and the 61st on economic ones, which a Western left--right instrument would read as consensus on one axis and middling disagreement on the other, yet on democratic commitment it lies at the 88th, on religion in government at the 88th, and on anti-elite sentiment at the 80th. Rather than asking whether societies' polarization looks like polarization in the United States, this paper asks what each society is polarized about.

Few scholars contend that politics is one-dimensional \citep{feldman2014understanding,carmines2015new}. The single ideological axis is inherited from studies of the 1970s U.S. Congress \citep{poole1985spatial,clinton2004statistical}, where it shows remarkable predictive power. It has survived as the default because measuring many dimensions of political attitudes globally requires lengthy survey batteries shared across many countries, which no survey program supplies. Regional barometers ask at length about democracy, corruption, and trust in institutions but say little about redistribution \citep{data_afro_r5,data_afro_r6,data_afro_r7,data_afro_r8,data_afro_r9,data_afro_r10,data_arab_w1,data_arab_w2,data_arab_w3,data_arab_w4,data_arab_w5,data_arab_w6,data_arab_w7,data_arab_w8,data_asian_w4,data_asian_w5,data_asian_w6}; the European Social Survey and the International Social Survey Programme investigate redistribution and social values but cover a fraction of the world \citep{data_ess1,data_ess2,data_ess3,data_ess4,data_ess5,data_ess6,data_ess7,data_ess8,data_ess9,data_ess10,data_ess11,data_issp2006,data_issp2008,data_issp2009,data_issp2010,data_issp2013,data_issp2016,data_issp2018,data_issp2019,data_issp2020,data_issp2023}; and national election studies ask a wide battery of unfortunately incomparable questions in one country each \citep{data_anes2024}. Prior work has either harmonized these surveys \citep{sdr2023,kolczynska2022combining} or fitted measurement models to country aggregates \citep{caughey2019policy,claassen2019estimating,berwick2025modgirt}, to a single program \citep{allison2021religiosity} or to parties rated by experts \citep{delacerda2026global}, leaving a critical gap in estimating preferences at the individual level worldwide (Appendix~A). The methodological obstacle is what the scaling literature calls a bridging problem \citep{bailey2007comparable,shor2010bridge,jessee2016estimate}---an Afrobarometer respondent and a European Social Survey respondent answer disjoint questionnaires---and it can be solved by either respondents or survey questions shared across instruments, or more recently, with synthetic respondents \citep{kaufman2026bridge}; we use such respondents only to set the direction in which each item runs before the model is fitted (Methods).

Here, we harmonize 1,378 items from twelve cross-national survey programs and national election studies in 17 countries into a common item bank and fit a measurement model in which every survey item maps onto the same underlying scales, so that people who answered different questionnaires can be compared directly. Shared survey items, identified by hand, become bridges so that all political positions are measured in the same standardized units. Drawing on the education literature, where researchers assembled incompatible international assessments into a single global learning scale \citep{angrist2021measuring}, we use bridging to place 2.13 million people interviewed in 155 countries since 2013 on five dimensions of political attitudes: social liberalism, economic left--right, democratic commitment, anti-elite sentiment, and religion in government. This improves on existing cross-national measures of polarization which rely on either party positions or on a single left--right self-placement item whose meaning varies across countries \citep{king2004enhancing,bauer2017leftright}; our latent positions estimated from a common item bank yield estimates of polarization, demographic disparities, and more, directly comparable across any country in our data set.

Formally, we study issue polarization, disagreement over political positions \citep{converse2006nature,knight2006transformations,jost2009political}, not the affective hostility between partisans \citep{gidron2020american,gidron2023who}, nor do we study polarization as the political process by which elites and institutions organize disagreement into opposing and mutually distrustful factions \citep{mccoy2018polarization,mccoy2019toward,svolik2019polarization}. We measure issue polarization as a public's dispersion of preferences \citep{dimaggio1996have,lelkes2016mass}. We also report three related analyses: (1) ideological constraint, or how different policy preferences cohere; (2) demographic cleavages and their relationship to political divides; and (3) whether the divisions we find among publics are the ones their parties are divided over.

Our central finding is that what a society is divided about varies. Two-thirds of European and Central Asian countries divide most over social liberalism; Middle Eastern and North African countries divide most often over democratic commitment, in 40\% of cases; Latin American countries divide most over economic left--right, in 59\%; and 24 of the 41 Sub-Saharan African countries we can measure divide most over anti-elite sentiment. East Asia has no single answer: social liberalism leads in 5 of its 17 countries, and economic left--right and religion in government in 4 each. Sixty-two of 154 countries are, like Iraq, most divided on democratic commitment, or anti-elite sentiment. Religion in government, the cleavage most associated with the Islamist--secular divide in the Middle East, is the sharpest division in ten countries, four of them in Europe and Central Asia, four in East Asia and the Pacific, and none in the Middle East. Yet, despite broad differences in what divides countries, we find that political disagreement is mostly domestic rather than reflecting fundamental differences across publics. On every dimension, 58 to 91\% of the variation in attitudes lies within countries rather than between them, and even on religion in government, the dimension closest to a clash of civilizations \citep{huntington1996clash}, most of the disagreement is internal.

Three further results illustrate how these divisions are organized. First, the dimensions rarely line up within people. The median within-country correlation between a person's social and economic positions is $0.01$, and the coherence \citet{converse2006nature} described is confined to the politically attentive voters in a small set of countries. Second, demographic groups divide weakly and inconsistently. We compare six pairs of groups within each country: old and young, more and less educated, richer and poorer, religious and non-religious, women and men, and urban and rural residents. Ideological gaps between these groups are small, typically less than a quarter of a country's internal spread. Nor do the groups diverge in the same direction everywhere. The highly educated are more anti-elite than those with only primary schooling in 36 of 39 Sub-Saharan African countries and 14 of 17 in East Asia and the Pacific, but less anti-elite in 33 of 44 European and Central Asian countries and in both the United States and Canada. Even the widening gap between young women and young men reverses in 35 of the 137 countries we can measure. Third, parties are often polarized over something other than what divides their publics. The dimension on which a country's parties are most polarized matches the one on which its public is most divided in 45\% of countries, about twice the 22\% expected by chance; across countries the two correlate most on social liberalism, and on anti-elite sentiment they correlate in opposite directions. Together, these findings suggest that the familiar Western pattern, in which a single left--right conflict organizes citizens' views, demographic divisions, and party competition at once, is the exception rather than the rule, and that comparative studies of polarization should begin by establishing what each society is divided about.

\section*{Data, harmonization, and the common scale}

To produce our comparative account of polarization, we assemble a single item bank from 99 waves of twelve cross-national survey programs and 23 national election study files from 17 countries: the World Values Survey \citep{data_wvs_ts,data_wvs6,data_wvs7} and the European Values Study \citep{data_evs_trend,data_evs_wvs_joint}, the Afrobarometer \citep{data_afro_r5,data_afro_r6,data_afro_r7,data_afro_r8,data_afro_r9,data_afro_r10}, the Arab Barometer \citep{data_arab_w1,data_arab_w2,data_arab_w3,data_arab_w4,data_arab_w5,data_arab_w6,data_arab_w7,data_arab_w8}, the Asian Barometer \citep{data_asian_w4,data_asian_w5,data_asian_w6}, the European Social Survey \citep{data_ess1,data_ess2,data_ess3,data_ess4,data_ess5,data_ess6,data_ess7,data_ess8,data_ess9,data_ess10,data_ess11}, the Latinobar\'{o}metro \citep{data_latino_ts,data_latino2023,data_latino2024}, the AmericasBarometer \citep{data_lapop_grand,data_lapop2023}, the International Social Survey Programme \citep{data_issp2006,data_issp2008,data_issp2009,data_issp2010,data_issp2013,data_issp2016,data_issp2018,data_issp2019,data_issp2020,data_issp2023}, the Comparative Study of Electoral Systems \citep{data_cses_imd,data_cses2,data_cses3,data_cses4,data_cses5,data_cses6}, the European Election Studies \citep{data_ees2009,data_ees2014,data_ees2019,data_ees2024}, the Pew Research Center's Global Attitudes Survey \citep{data_pew2022,data_pew2023,data_pew2024,data_pew2025}, and the national election studies of Australia, Austria, Britain, Canada, Denmark, Finland, Germany, Iceland, Ireland, Israel, Italy, the Netherlands, New Zealand, Norway, Switzerland, Taiwan, and the United States \citep{data_aes2022,data_autnes,data_bes_rps2024,data_bes_panel,data_ces2021,data_dnes,data_fnes2023,data_gles2025,data_icenes2021,data_ines2024_f2f,data_ines2024_online,data_ines2025_pres,data_ines_israel2022,data_itanes2022,data_dpes2023,data_dpes2023_migrant,data_nzes2023,data_nes_norway2021,data_selects2023,data_teds2025,data_anes2024}. The assembled corpus contains 122,620,321 respondent-item observations from 3,997,205 respondents in 161 countries, fielded between 1981 and 2025 (Table~\ref{tab:corpus}). We estimate the measurement model across the full corpus so that every survey wave helps locate the items' latent positions, but we confine the analyses that follow to the 2,131,072 respondents in 155 countries surveyed in 2013 or later so that the estimates describe contemporary publics rather than averaging over four decades of change.

Our harmonization procedure produces \textit{alignments}: two or more items from different instruments that measure the same construct. We recode an alignment's constituent items onto a common $[0,1]$ scale and orient them semantically. We construct alignments manually from question text and codebooks, validate each in two independent ways, and verify the orientation of every aligned item (see Methods). Of 1,378 survey items in our battery, 152 are alignments that appear in two or more programs, and the corpus is a single connected graph (Figure~\ref{fig:linkage}).

We measure political attitudes on five dimensions. Social liberalism captures acceptance or rejection of homosexuality, abortion, immigrants, and ethnic minorities, and economic left--right measures support for or opposition to redistribution, progressive taxation, and government responsibility for welfare and health care. Democratic commitment combines authoritarian values with satisfaction with how democracy is working, and anti-elite sentiment aggregates perceptions of corruption with institutional distrust. Religion in government measures support for religious authority over law, government and politics (rather than religiosity itself). We assign each harmonized item to one or two of these dimensions and leave 399 (29\%) unassigned. For each dimension we fit a graded response model \citep{samejima1969graded,chalmers2012mirt} to human responses across all surveys, removing respondents who answered fewer than two of that dimension's items; the fitted models contain 22 to 311 items and 319,058 to 3,483,383 respondents each, with marginal reliabilities of 0.83 for anti-elite sentiment, 0.74 for economic left--right, 0.70 for social liberalism, 0.68 for religion in government and 0.60 for democratic commitment (Table~\ref{tab:dims}). We detail the coverage and reliability tests for these five dimensions in Methods.

A sample of the results of this model is in Figure~\ref{fig:ridges}; the full results are in Appendix~C (Figure~\ref{fig:ridgesall}). Each row shows one of the countries in our sample, colored according to its region; each column contains countries' distribution of preferences for a given dimension. Rows are ordered by mean social liberalism score. In the first row, Andorra is the most socially liberal country, but much more moderate economically. Too few Andorrans answer three or more religion items for a distribution to be drawn, so the fifth column is empty.

\begin{figure}[htbp]
\centering
\includegraphics[width=\linewidth]{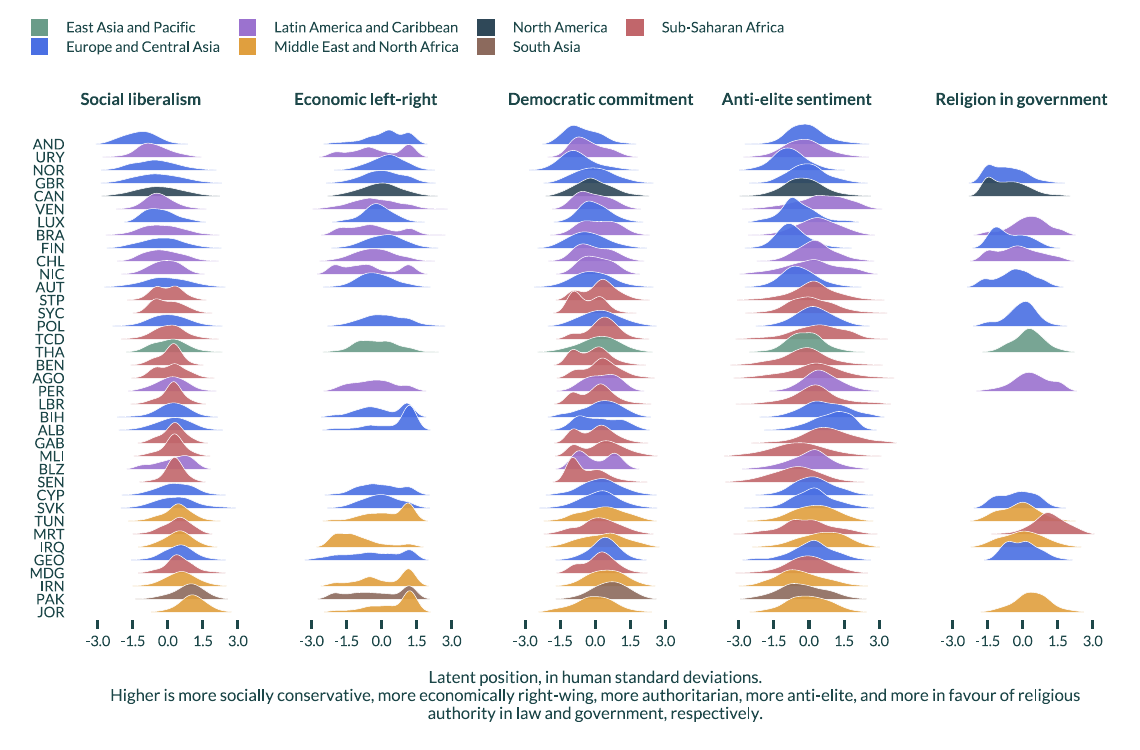}
\caption{\textbf{The political opinion of the world, by country and dimension.} Each ridge is the distribution of one country's respondents on one dimension, in human standard deviations; higher values are more conservative, authoritarian, anti-elite, or favorable to religious authority. Countries are ordered by mean social liberalism in every panel, and colored by region. Not every country can be measured on every dimension.}
\label{fig:ridges}
\end{figure}

\section*{Political disagreement is mostly domestic}

Whether the world's political conflict exists between nations or within them is an old question, bearing implications for international conflict and cooperation. One tradition treats countries or civilizations as coherent political communities whose publics differ systematically from each other \citep{huntington1996clash,inglehart2005modernization}; another locates the largest conflicts within societies \citep{norris2019cultural}. Our individual-level estimates on a common scale can begin to answer this question directly. On every dimension, most of the variation in attitudes around the world lies within countries rather than across them: 91\% on economic left--right, 78\% on anti-elite sentiment, 70\% on democratic commitment, and 62\% on social liberalism, after removing measurement error and differences between survey programs (Appendix~C). Economic preferences such as patience and risk-taking show the same pattern across 76 countries \citep{falk2018global}. The exception is the dimension closest to the civilizational account, religion in government, on which countries differ more than on any other. Yet, even there, 58\% of the variation lies within them. Nor does a country's average position on an attitude indicate how divided it is: across the five dimensions, the mean and the dispersion correlate between $-0.23$ and $+0.24$. A country's average attitude is seldom indicative of how deeply its citizens disagree with one another. This is the question motivating the remainder of this paper.

\section*{What countries are divided about}

Our first research interest is to identify which dimension of political conflict most divides each country. However, there are important measurement challenges to comparing the dispersion of attitudes across different dimensions and countries. First, each dimension has an arbitrary unit scale. Second, not every country can be measured on all five dimensions. We resolve both issues by modeling each country's spread on each dimension as the sum of a dimension effect, which captures the scale's unit, and a country effect, which controls for the overall level of division within that country. Under this measure, a country's sharpest division is the dimension on which its spread most exceeds what those two effects predict. Appendix~E compares this choice with two others that fix the units differently, ranking countries within each dimension, and assuming equal worldwide variance on every dimension; 98 of the 154 countries have the same sharpest division under all three.

\begin{figure}[htbp]
\centering
\includegraphics[width=\linewidth]{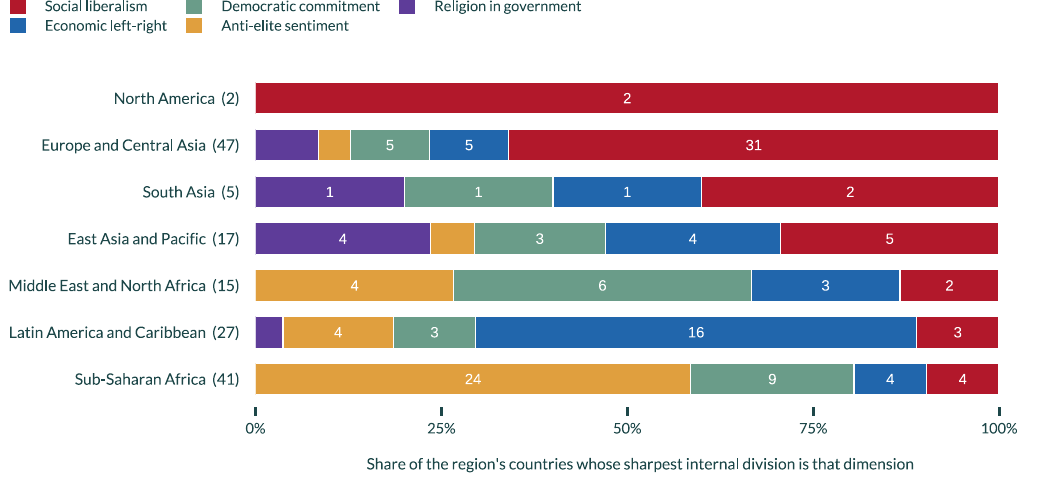}
\caption{\textbf{Most divisive political dimension by region.} Share of each region's countries whose sharpest division is each dimension, among 154 countries measured on at least two dimensions. The sharpest division is the dimension on which a country's spread most exceeds the prediction of a two-way model with dimension and country effects (Methods; Appendix~E). Regions follow the World Bank classification.}
\label{fig:whichdimension}
\end{figure}

Figure~\ref{fig:whichdimension} shows that, overall, no dimension dominates across the 154 countries: social liberalism is the sharpest division in 49, anti-elite sentiment in 35, economic left--right in 33, democratic commitment in 27, and religion in government in 10. However, there are distinct regional patterns. In Europe and Central Asia, 31 of the 47 countries (66\%) are divided most over social liberalism. In the Middle East and North Africa the modal division is democratic commitment, in 6 of 15 countries (40\%); in Latin America it is economic left--right, in 16 of 27 (59\%); and in Sub-Saharan Africa 24 of 41 countries divide most over anti-elite sentiment. East Asia and the Pacific has no dominant division, with social liberalism leading in 5 of 17 countries and economic left--right and religion in government in 4 each. Among the 98 countries with the same sharpest division under all three rules, 17 European countries divide most over social liberalism, 14 Latin American countries over economic left--right, and 24 African countries over anti-elite sentiment. These results are also robust to estimation error: under a bootstrap approach, a country's sharpest division is unchanged in at least 95\% of draws for 132 of the 154 countries. Furthermore, among the 85 countries measured on all five dimensions, social liberalism is the sharpest division in 33, economic left--right in 23, democratic commitment in 14, religion in government in 10, and anti-elite sentiment in 5, and the modal dimension is unchanged in Europe, Latin America, and the Middle East; only 5 of the 41 African countries are among them.

While religion in government divides publics in the Middle East, it is never what divides them most. To compare countries on a single dimension we use each country's percentile of spread on it. The Islamist--secular cleavage is widely studied as a Middle Eastern phenomenon \citep{ciftci2013secular,wegner2019revisiting}, and the region's publics are split over it: Iran, Iraq, Yemen, and Lebanon sit at the 89th, 88th, 86th, and 83rd percentiles of religion spread. In none of them, though, is it the sharpest division: economic left--right divides Iran more, democratic commitment divides Iraq and Lebanon more, and anti-elite sentiment divides Yemen more. Of the 10 countries where religion in government is the sharpest division, four are in Europe and Central Asia, four in East Asia and the Pacific, and one each in South Asia and Latin America. In six of them it is a large division by any standard, above the 90th percentile of religion spread: Puerto Rico, Romania, New Zealand, Cyprus, Ukraine, and the Netherlands. Note that these comparisons rely on the 93 countries where religion in government is measured; of these, 35 are in Europe and Central Asia and 13 in the Middle East and North Africa. As such, this comparison describes where religion's role is contested among the places where survey batteries ask about it.

Note also that our data are weakest in Africa. Most African countries are measured on two or three dimensions only, so while anti-elite sentiment appears most divisive across the continent, we can measure fewer dimensions that might compete with it (Methods). Worldwide, anti-elite sentiment is the sharpest division in 27 of the 42 countries measured on two or three dimensions but only in 5 of the 85 measured on all five, so we cannot definitively separate an African pattern from the composition of the comparison. Our results for Africa should therefore be taken as suggestive.

\section*{Ideological constraint is rare}

Our second research question asks whether and under what circumstances do individuals' preferences across these scales correlate. This question draws from \citet{converse2006nature}, who argued first that most people do not organize their opinions according to an ideological system, and that those who do are the ones who follow politics most closely. Both claims have been tested extensively in the United States \citep{ansolabehere2006purple,ansolabehere2008strength} and, more recently, across European democracies \citep{warncke2025what}, but not yet across the world's regions and many opinions at once.

Overall, the median within-country correlation between social liberalism and economic left--right is $0.01$ across 96 countries, and it is positive in only 52\% of them. This agrees with the one earlier test at comparable breadth, which found across ninety-nine nations that cultural and economic attitudes are rarely organized along a right--left line \citep{malka2019cultural}. The other nine pairs of dimensions behave similarly, with medians between $-0.11$ and $+0.18$. The largest is religion in government with the authoritarian pole of democratic commitment (median $+0.18$, 42 countries); religion in government is, however, unrelated to social liberalism ($0.00$, 29 countries).

While the median within-country correlation between social and economic scales is near 0, it is occasionally much higher, most of all in North America and parts of Western Europe: $+0.54$ in the United States, $+0.48$ in Canada, $+0.37$ in the Netherlands, and $+0.27$ in Switzerland and Australia, followed by Austria ($+0.25$) and Sweden ($+0.18$); outside the West it reaches $+0.19$ to $+0.29$ in Pakistan, Tajikistan, Bangladesh, Tunisia, and Malaysia, and $+0.11$ to $+0.16$ in Kenya, India, Turkey, Jordan, and Azerbaijan. Conversely, in seven countries the correlation is below $-0.10$, most of all in Hungary ($-0.16$).

\begin{figure}[htbp]
\centering
\includegraphics[width=\linewidth]{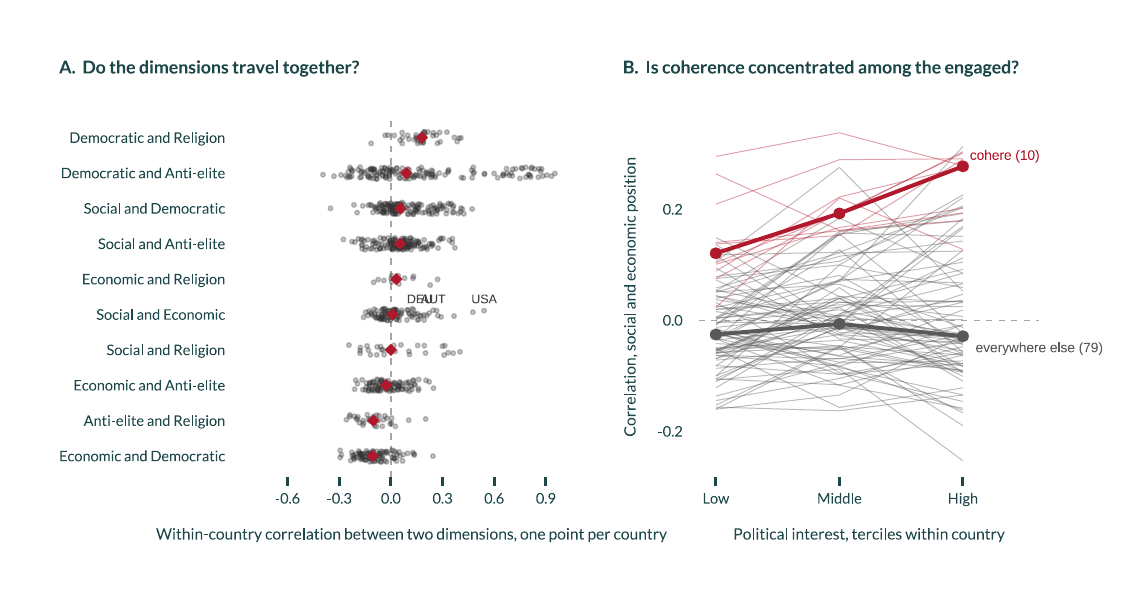}
\caption{\textbf{Ideology coheres in a small and specific set of countries, and there it is concentrated among the attentive.} \textbf{A}, within-country correlation between each pair of dimensions: one gray point per country, red diamonds at the median. \textbf{B}, the social--economic correlation within each tercile of political interest: thin lines are the 89 countries with at least 100 respondents per tercile, bold lines the medians. The ten countries in red are those whose correlation exceeds $+0.15$ in a random half of their respondents with a political-interest measure (Appendix~F).}
\label{fig:coherence}
\end{figure}

Among respondents whose surveys ask about political interest, we identify ten countries where social liberalism and economic left--right cohere, with a correlation above $+0.15$ in a random half of each country's respondents, and measure the gradient on the other half (Appendix~F). In those ten, the correlation rises from $+0.12$ among the least politically interested tercile to $+0.28$ among the most; in the remaining countries, the dimensions are uncorrelated at every level of political engagement. Attention also does not induce ideological constraint along a country's own sharpest divisions: the two dimensions a country is most divided on correlate at a median of $+0.16$ across 71 countries, but no more among its most interested citizens than among its least (Appendix~F). These results suggest that, for most of the world, a person's position on one dimension carries essentially no information about their position on another. Converse's two observations, that most citizens hold weakly connected positions and that constraint rises with political attention, both reappear, but the second only for the social--economic dimensions and only among the countries where those dimensions correlate well (Figure~\ref{fig:coherence}).

\section*{Demographic divisions are small and unstable}

In established Western democracies, political conflict is routinely interpreted through a lens of social groups. The cleavages \citet{lipset1967cleavage} identified, between center and periphery, church and state, land and industry, and owners and workers, organized party systems for most of the twentieth century, and recent accounts swap class for education, age, and urban or rural status. University graduates have moved toward parties of the left while voters without degrees have moved toward the right \citep{gethin2022brahmin}; older cohorts anchor the conservative side of debates over immigration and gender \citep{norris2019cultural}; and the gap between cities and the countryside in how people vote has widened sharply in the United States, Britain and Canada and has opened in several European multiparty systems \citep{rodden2019why,huijsmans2025great}.

Our third research question asks two things of these groups. Do they divide on every dimension of political conflict, or only on some: are older citizens as far from younger ones on democratic commitment and anti-elite sentiment as on social liberalism? And do they divide the same way everywhere: are older citizens more socially conservative in every country, or only in Western democracies? Broadly, the groups fail the second test: for most combinations of group and dimension, the gap points in opposite directions in different countries.

\begin{figure}[htbp]
\centering
\includegraphics[width=\linewidth]{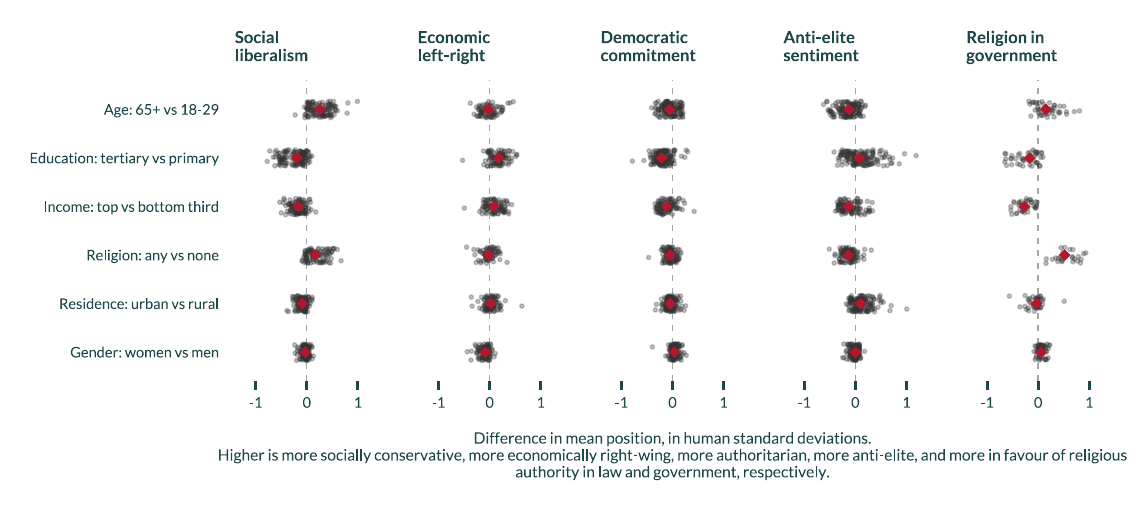}
\caption{\textbf{Demographic divisions sort political opinion weakly, and not in the same direction everywhere.} Each gray point is one country's gap between the two named groups on one dimension, in human standard deviations; red diamonds are medians across countries.}
\label{fig:cleavages}
\end{figure}

Answering either question requires individual-level ideology estimates paired with harmonized demographic variables, which our data set is the first to so extensively supply. We compare six pairs of groups: respondents aged 65 and over against those aged 18 to 29, the tertiary-educated against those with only primary schooling, the religiously affiliated against those with no religion, the top third of a country's income distribution against the bottom third, urban against rural residents, and women against men. For each pair, dimension, and country, we calculate the gap between the two groups' mean positions, in human standard deviations, wherever both groups have at least 100 scored respondents, and summarize each pair and dimension by the median gap across countries (Figure~\ref{fig:cleavages}). Respondents over 65 are 0.25 standard deviations to the socially conservative side of those under 30, the religious 0.17 to the conservative side of the non-religious, the top third of earners 0.09 to the economic right of the bottom third, and the tertiary-educated 0.21 toward the democratic pole of democratic commitment. Leaving aside religious affiliation's relationship with religion in government, the largest median gap is 0.27 standard deviations, between the top and bottom thirds of earners on religion in government, against a median within-country spread of about 0.85. The urban-rural divide, a stark feature of the contemporary political landscape in the United States, is among the weakest cleavages, with a median gap of 0.06 to 0.11, although our data show that the United States has the largest urban--rural gap in the corpus on social liberalism and the third largest, after Libya and Georgia, on economic left--right.

However, these divisions do not point the same way everywhere. In the typical pairing of a group with a dimension, the gap runs opposite to the median in about one country in four. Only six of the 30 pairings point the same way in at least nine countries in ten. Older respondents are more socially conservative; the educated are more socially liberal and more committed to democracy; the rich are more socially liberal; and, in the few countries where religion in government can be measured, the religious favor religious authority more and the rich less. Even the most well-documented demographic cleavage is not universal: the old are more socially conservative than the young in 95\% of 120 countries, and more liberal in the other six: by 0.23 standard deviations in Israel and by less than 0.05 in Nigeria, Indonesia, Mali, Guinea, and Uganda.

Lastly, a generational gender gap, with younger women to the left of younger men, has been documented for decades in postindustrial societies \citep{inglehart2000developmental}. The gender gap in social liberalism widens steadily from older to younger cohorts: among respondents over 65 it is close to zero, with women the more liberal in 52\% of countries, while among 18-to-29-year-olds women are the more liberal in 74\% of 137 countries. In the other 35, among them Bangladesh, Kyrgyzstan, Colombia, Thailand, and Ireland, young women are more conservative (Appendix~D); the gap exists well beyond the societies in which it was first described, but it is not universal.

\section*{Parties often divide over something other than what divides their publics}

What the public disagrees about most need not be the same issues that divide parties; party systems can remain organized around older cleavages after the divisions within the electorate have moved on \citep{lipset1967cleavage}. Our fourth research question asks how often the public's most divisive dimension aligns with parties'. To answer it, we use a measure of party-system polarization from V-Party, which places the parties of 178 countries on expert-coded scales \citep{lindberg2022vparty,dupont2022global}. We give each party a position on each of our five dimensions, averaging the two V-Party indicators that match it (religion in government has one). A party system's polarization on a dimension is then the spread of its parties' positions at the country's most recent election between 2010 and 2019, weighting each party by its vote share \citep{dalton2008quantity}. We identify the dimension that divides a country's parties most with the same two-way model we use for publics (see Methods). 

We compare publics with parties in three ways. First, we ask whether a public's sharpest division is also its parties'. Second, for each dimension, we ask whether countries whose publics are more divided on it have more polarized parties on it. Third, we ask where the dimension that divides a country's parties most falls in its public's own ranking.

In the 121 countries with both measures, the dimension that divides the parties most is the one that divides the public most in 45\%, against 22\% expected by chance, and in 31 to 42\% under the alternative ways of combining the indicators (Appendix~H). These public-party matches are concentrated where publics divide most over social liberalism (27 of 47 countries) or democratic commitment (11 of 20); where they divide most over economic left--right or, intuitively, anti-elite sentiment, their parties more often than not polarize over something else (they match in 3 of 23 and 10 of 22).

\begin{figure}[htbp]
\centering
\includegraphics[width=\linewidth]{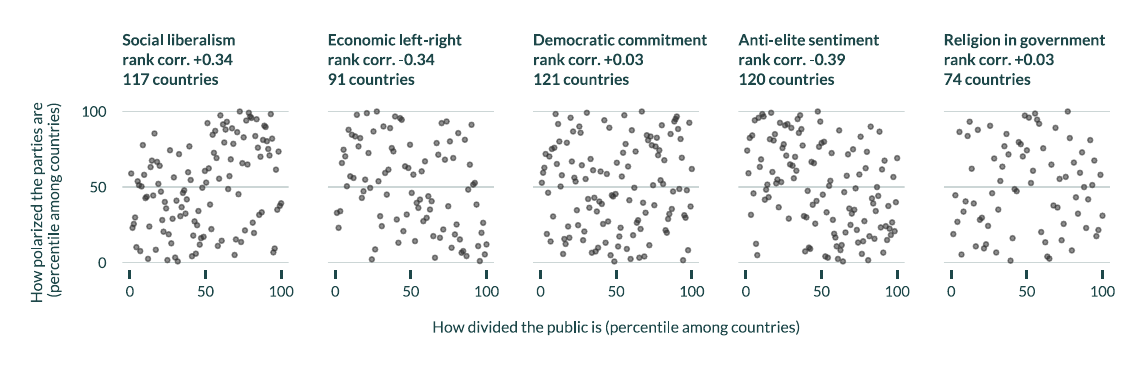}
\caption{\textbf{Parties and publics divide on different issues.} Each point is a country. The horizontal axis gives how divided its public is on the dimension: the percentile of its spread among the countries measured on that dimension. The vertical axis gives how polarized its party system is on the same dimension: the percentile of the vote-weighted standard deviation of party positions at its most recent V-Party election between 2010 and 2019, with two V-Party indicators averaged into one position per party (one for religion in government; Methods). Panel titles give the Spearman rank correlation and the number of countries.}
\label{fig:partypublic}
\end{figure}

In our second analysis, we find that public and party polarization correlate positively only on social liberalism (rank correlation $+0.34$ across 117 countries; Figure~\ref{fig:partypublic}). On democratic commitment, the two are unrelated ($+0.03$), even though in 11 of the 20 countries whose publics divide most over it the parties polarize over it too, and they are unrelated on religion in government ($+0.03$ across the 74 countries where religion spread and party positions both exist). On economic left--right and anti-elite sentiment, they correlate negatively ($-0.34$ across 91 countries and $-0.39$ across 120): the publics most divided over elites have the party systems least polarized on anti-elitism. 

Third, when parties and publics do not share their sharpest division, the parties often choose a dimension of secondary importance. The dimension that divides a country's parties most is its public's sharpest division in 45\% of countries, and its second-sharpest in a further 21\%; it is the public's least divisive dimension in only 17\%. Some party divisions, however, lie outside the attitudes we measure. In Iraq and Lebanon the Sunni--Shia divide predicts which party a respondent feels closest to almost perfectly (Cram\'{e}r's $V$ of 0.67 and 0.97) while explaining under 1\% of the variance on democratic commitment. 

Two further measures in Appendix~H, how strongly party supporters are sorted by their views and which issues publics name as most important, agree on social liberalism: there, wider public disagreement goes with more polarized parties, more sorted partisans, and greater issue importance. On no other dimension do all three line up; on anti-elite sentiment, partisans are sorted most where the public is most divided, largely along the line between government and opposition, even though party positions are not more polarized there.

\section*{Discussion}

This study places 2.13 million people surveyed in 155 countries since 2013 on five dimensions of political attitudes and asks what each society is divided about. National publics differ sharply in where their disagreement lies according to regional patterns: social liberalism in Europe, democratic commitment in the Middle East and North Africa, anti-elite sentiment in Africa, economic left--right in Latin America, and religion in government, where it can be measured, in a scattering of countries across Europe and East Asia. Secondary results qualify how that disagreement is organized. Positions on different dimensions rarely line up within people; demographic groups divide weakly and in different directions in different countries; and parties are most polarized over what their publics are most divided over in fewer than half of countries, with party polarization tracking public division across countries only on social liberalism. Overall, we find that while a wide variance shows that people disagree, it does not imply that their politics is organized around that disagreement.

The results carry three implications for how polarization is studied. First, a single left--right index does not capture the extent of polarization in most countries outside of Western democracies. Second, a country's average position on a political dimension, the quantity measured and reported in most cross-national studies, says little about how divided that country is (across our five dimensions, mean and spread correlate between $-0.23$ and $+0.24$; Appendix~G); studies of polarization's causes and consequences should consider distributions rather than only means. Third, the familiar account in which attentive citizens sort into consistent left--right camps and parties compete over the divisions their voters hold describes ten countries, only half of which are wealthy democracies (Appendix~F). Elsewhere, attention does not bundle positions, demographic groups do not reliably divide, and parties often contest something other than what their publics disagree about most, which leaves wide disagreements, over elites above all, that party competition does not capture. Theories of representation and democratic backsliding should therefore test whether the Western pattern holds before presupposing it.

We note a number of limitations in our measurement. The central one is a ubiquitous assumption in the scaling literature: that a given answer means the same thing everywhere \citep{king2004enhancing,davidov2014measurement,stegmueller2011apples,vandermeer2019putting}. We probe this assumption with regional refits, agreement between programs surveying the same country, and an item-level screen, and find broad robustness (Methods; Appendix~G). Dropping every item that fails the screen where it fails leaves country spread rankings correlated with ours at 0.997 or above and the most divisive dimension unchanged in 143 of 154 countries. Democratic commitment and economic left--right fare worst, chiefly between Africa and other regions, so where African publics are compared with others we read those two dimensions as orderings rather than distances. However, this is no substitute for more rigorous surveys that incorporate anchoring vignettes, broader survey batteries, and harmonized demographics, which we urge cross-national survey programs to adopt.

Four further limits bound our claims. First, we measure issue polarization, not the affective hostility between camps that much recent comparative work studies \citep{gidron2020american,gidron2023who}. Second, what we observe as dividing a country most is at least partly a property of the questions it was asked. Where two programs survey the same country, its profile of spreads across dimensions agrees at a mean correlation of only 0.17 (95\% interval 0.05 to 0.28, 75 countries), against 0.49 between two waves of the same program, and the two name the same most divisive dimension in 41\% of cases against 33\% by chance (Appendix~E). Since most regions are covered most comprehensively by their own barometer, the regional pattern partly reflects which instrument surveys each region, and a country's most divisive dimension describes its public as its surveys measure it rather than a fixed property of that public. We do find that party alignment, the share of a dimension's variance that party support explains, generalizes better across programs and points to a partly overlapping regional pattern (Appendix~H). Third, the scale's largest block of questions, the national election studies, bridges to the cross-national programs' economic dimension through only five items (Table~\ref{tab:linkage}), which limits the conclusions economic left--right can support. 

Fourth, and perhaps most importantly, our results cannot speak to the question of widest interest in this literature: whether polarization is increasing over time. All our estimates are cross-sectional, a snapshot of the early 2020s assembled from sources fielded between 2013 and 2025. This is a limitation of our corpus rather than our method, and one we expect we can overcome as more waves of barometers and national election studies are released. To facilitate this analysis, we release the item bank and the fitted model parameters for all five dimensions; because the parameters are held fixed, a respondent surveyed in 2030 can be trivially scored onto the scale we estimate today once their survey's items are harmonized to our question bank. 

In this paper we have built what is, to our knowledge, the first set of individual-level political attitude estimates on a common scale at this breadth: 1,378 harmonized items from 99 waves of twelve cross-national programs and 23 national election studies, 4 million respondents in 161 countries interviewed between 1981 and 2025, placed on five dimensions by replicable measurement models, all of which we make publicly available (see Data and Code Availability). Because the estimates are individual rather than national, they can be joined to everything else a survey records about its respondents, and we hope that makes them useful well beyond the study of polarization. Research on democratic support and backsliding can ask which and how many citizens would sacrifice democracy for policy gains, in comparable units across countries \citep{graham2020democracy,svolik2019polarization}. Research on representation, which already compares the full distribution of a public's views with its parties' \citep{golder2010ideological,lupu2017new}, can do so in common units across countries and against policy outcomes \citep{lindberg2022vparty}. Research on migration and on religion and politics can ideologically locate a diaspora between its country of origin and its country of residence. The same data offer a benchmark for silicon sampling, in which language models prompted with a persona stand in for survey respondents \citep{argyle2023out}: a simulated public in any of 155 countries can be checked against the full distribution of the real one, by demographic group and on five dimensions, rather than against a handful of national averages \citep{santurkar2023whose,durmus2024globalopinions}.

\section*{Ethics statement}

This study analyzes secondary data from public survey archives. No human subjects were recruited and no individually identifying information was accessed: every survey file used here is a de-identified public release distributed by its program, used under that program's terms.

Estimating positions for individual respondents rather than for countries raises a disclosure question that a country-level design does not, and it resolves in this case because we add no information about any person. Every variable we use --- the survey responses, and the demographic and political covariates joined to them --- is already public, distributed in that form by the originating program. The latent estimates are a function of those public responses and the fitted item parameters, keyed to the pseudonymous respondent identifiers the releases already carry. A respondent's estimate is therefore recoverable by anyone holding the source file and identifies no one to anyone without it, the same standing the underlying responses have had since the programs published them.

Two further considerations bear on how the results should be used. Describing a population or a subgroup as distant from some median is a measurement claim and not a normative one. We take no position on whether any of the political attitudes measured here are correct, and nothing in this analysis implies that agreement with any median is desirable; a country reported here as socially conservative relative to the world median is being located, not judged. The countries where our estimates rest on the fewest items are also concentrated in the global South, where survey infrastructure is thinnest. Caution about specific numbers is therefore greatest there, and the marginal survey item would contribute most there as well, which is a reason to report coverage in full rather than to restrict the analysis to the places already well served.

\section*{Data and Code Availability}

We obtain all survey data from their originating programs and redistribute them subject to each program's license. Every release we use is cited where its program is introduced, with its version and, where the archive issues one, its persistent identifier. The replication archive contains the harmonization code, the alignment map, the item bank, the fitted item parameters, the individual-level latent estimates, the covariate extraction and harmonization code, and the language-model personas' prompts, raw replies, and decoded answers. We release the fitted item parameters alongside the estimates, allowing a survey we have not seen to be placed on the scale without refitting. The code comes in two parts: a corpus build, which turns the raw source files into the harmonized corpus and rests on hand-coded harmonization and orientation decisions released with it, and a numbered analysis pipeline that runs from the corpus to every figure and number reported here.

\section*{Author Contributions}

A.O. produced the first version of the data set, as part of his undergraduate thesis. A.R.K. conceived the study, produced the final version of the data, carried out all analysis, developed the framing, and wrote the first draft. H.I. revised the figures and draft and contributed to the framing.

\section*{Competing Interests}
The authors declare no competing interests.

\section*{Acknowledgments}

We thank Chris Barrie, Noam Gidron, Yusaku Horiuchi, Alex Siegel, Jonathan Rodden, Yamil Velez, and seminar participants at the University of Tokyo for helpful feedback and discussions. This work is supported in part by the NYUAD Center for Interdisciplinary Data Science \& AI (CIDSAI), funded by Tamkeen under the NYUAD Research Institute Award CG016.

\clearpage

\newpage

\bibliographystyle{unsrtnat}
\bibliography{main}

\clearpage
\section*{Methods}

The main text describes the corpus and the five latent dimensions, and Appendix~B the validation of the common scale against held-out self-placement. This section describes the harmonization procedure and its checks, the estimation specifications and thresholds, the covariate construction, and the screen for differential item functioning.

\subsection*{Harmonization checks and linkage}

Harmonization proceeds by grouping items into alignments. An alignment is a claim that two or more items from different instruments measure the same construct; we recode its members onto a common $[0,1]$ scale, orient them semantically, and treat them as a single column. We construct alignments manually from question text and codebooks rather than by string similarity, because substantive equivalence and lexical similarity frequently diverge: ``immigrants take jobs away'' and ``immigrants are good for the economy'' share almost no vocabulary yet measure the same construct in opposite directions.

We validate each alignment in two ways. First, the members of an alignment must agree at the country level, since two versions of the same question that correlate negatively across countries cannot both measure the construct in the stated direction. Second, items within a dimension must agree with the mean of all other items in that dimension, so that no single item's orientation determines the outcome. We treat failures of the first check as defects and repair them; failures of the second may indicate that a dimension captures more than one trait, which is the case for two of our five dimensions and is described below.

Item orientation is the step at which errors are most consequential, because a reversed item produces plausible values rather than a detectable failure; pooled with its counterparts, it attenuates their shared signal and inflates measured inconsistency for every respondent who answered it. Reversing 237 source items, of 286 whose orientation we ruled on individually, changed our estimates more than any modeling choice in the pipeline; the coding procedure and its validation are described below.

\paragraph{Linkage across programs.} A joint scale depends on the items connecting its parts; we therefore measure linkage from the assembled corpus itself rather than from the alignment map. Of 1,378 harmonized items, 152 (11.0\%) appear in two or more survey programs, and these 152 items carry the entire linkage. Treating the national election studies, with the CSES fielded inside them, as one block, 50 of the 66 pairs of programs share at least one item. The 16 that share none all involve the European Election Studies or the Pew Global Attitudes Survey, which join the rest through three items and two respectively; every program is nonetheless reachable from every other, and the corpus forms a single connected graph (Figure~\ref{fig:linkage}).

The linkage is not uniform, and its weakest points warrant description. The two newest programs are the thinnest joints in count: the Pew survey, which supplies five of the religion-in-government items, shares one item with the ISSP and one with the election studies, and the European Election Studies one each with the Asian Barometer, the European Social Survey, and the election studies. The largest block is the national election studies, which contribute 766 items and 424,376 respondents across 51 countries, counting the three CSES modules filed with them, and connect to the cross-national programs through 40 items, 32 of them on one of the five dimensions: seven on democratic commitment, thirteen on anti-elite sentiment, nine on social liberalism, five on economic left--right, and three on religion in government, with several items carrying more than one dimension. This asymmetry is a property of the instruments rather than of the harmonization. Establishing that an item runs in the same direction in two sources requires correlating country means, which in turn requires several countries in common, and every national election study file in the corpus covers a single country, the only multi-country studies in that block being the three CSES modules. For those, we establish direction manually from the questionnaire, and we keep items lacking either form of evidence as separate columns rather than pooling them. Hand-verified direction is scarcest for economic left--right, and its five linking items limit the conclusions that dimension can support.

\subsection*{Coding item direction}

An alignment is only usable if its members are known to run in the same direction, and direction is the part of harmonization that neither string matching nor covariance can settle. It depends on the valence of the statement rather than on the order of the options, which is the general form of the immigration example given earlier; two items can need opposite codings from identical option lists. Two mechanical procedures failed on that point, one reading option labels by rule (6 of 23 correct) and one correlating each item against its own study's dimension-mates (29\%, invalid because items within a study share an un-oriented raw coding).

\paragraph{Language-model assistance, and its validation.} We perform part of the remaining hand verification with a language model, as follows. We pose the question to Claude Opus 5, giving it the item text, the response options, and the target column's meaning, and nothing else.

We validate the procedure before use. Run against 56 items already coded by hand, it agrees on 52 (93\%), and on 48 of the 50 answers it marks high-confidence (96\%); two of the four disagreements are ones it had itself flagged as uncertain. We adopt only high-confidence answers, covering 22 of the 52 open items, and leave the remainder uncoded and therefore unpooled, which is the conservative outcome rather than a missing one. In a separate blind check we show the model 18 items whose hand coding an independent covariance-based screen disputes, without telling it which set any item belongs to or what the existing code is, and it agrees with the hand coding on 16. As with the denomination coding described below, the question put to the model is semantic rather than political --- what a scale's wording implies about its direction --- and we check the answer against human coding rather than assume it. We consult no model about any item's substantive content, its dimension, or its relationship to any political position.

\paragraph{A per-source direction screen.} The two checks described earlier are a cross-country check, which requires several countries and so cannot run on a single-country election study, and a check against the other items in a dimension, which is global. Neither catches a reversal confined to one national questionnaire, which is where a reversal is most likely to have survived. We therefore screen every item within every source: for each source, item and dimension with at least 300 responses we correlate the response category against the respondent's estimate on that dimension and compare the sign to that of the item's fitted discrimination. Agreement is the expectation, since the model absorbs an item running against its dimension as a negative discrimination. Of 2,008 such cells the signs disagree in 31, or 1.5\%; by dimension the rate is 1.2\% on economic left--right, none on social liberalism, 1.8\% on anti-elite sentiment and 2.7\% on democratic commitment. Thirteen of the 31 concern items whose fitted discrimination is below 0.15 in absolute value, which is to say items with no meaningful sign to disagree with, leaving 18 candidates that we release as a review sheet with question text attached. Each is a candidate defect rather than a confirmed one, since the estimate is built partly from the item being screened and an item carrying two traits will disagree in some sources without anyone having mis-coded anything; we report the screen as a bound on how much of this class can remain. The screen has one blind spot, and it is the case that matters most. Because each item is compared with an estimate that its own source's items help define, a source whose items on a dimension are reversed together agrees with itself and passes. That failure is invisible from inside the human data, and it is what the language-model personas of Appendix~B detect.

\subsection*{Fitting the dimensions}

The fits use \texttt{mirt} \citep{chalmers2012mirt}, with responses binned as $\min(\max(\text{round}(v(K-1))+1, 1), K)$, where $v$ is the harmonized $[0,1]$ value and $K$ the number of categories. Fourteen items from the 2024 American National Election Study are left out of every fit: they are strength-of-feeling follow-ups (``strongly or not strongly''), whose direction is set by the answer to the question before them, so on their own they carry no sign. Calibration uses every respondent in every wave, 1981 to 2025, answering at least two of that dimension's items rather than a capped draw; the resulting samples run from 319{,}058 to 3{,}483{,}383 respondents over 22 to 311 items, and all five fits converged (Table~\ref{tab:dims}). Those are the samples used to estimate the item parameters and not the people the analyses describe. Every respondent interviewed in 2013 or later and answering at least two items is scored against the parameters, held fixed, which places 2,131,072 people in 155 countries on at least one dimension; earlier waves inform the item parameters and enter no analysis. A country needs 100 scored respondents for a position averaged from individual estimates, a minimum every country with any scored respondent clears; the countries with none, 26 on economic left--right and 46 on religion in government, whose respondents answer a single item, are placed by the pooled-likelihood estimator alone (Appendix~C).

Some survey waves share no item with the rest of a dimension, or only one or two, and the model can fit such a block either way up with the same or nearly the same likelihood; estimation keeps whichever way it starts. Before fitting, we therefore code each item to run toward its dimension's pole, as judged by the answers of 75 language-model personas arranged from one pole to the other \citep{kaufman2026bridge}, and convert the parameters back afterwards, which reparameterizes the model rather than changing it; the personas enter no fit and no analysis (Appendix~B). On social liberalism, orienting the items raised the log-likelihood by 906 on the same model and data and set the reversed Dutch and German election-study blocks the right way up.

\subsection*{Dimension assignment and coverage}

We assign items to dimensions by a membership mapping that is not a partition; an item may load on more than one dimension, and 399 items remain unassigned and enter no analysis. The dimension labels can be read from the country orderings they produce. Social liberalism orders countries with Andorra, Iceland, New Zealand, Sweden, and the United States at the liberal pole and Jordan, Libya, Kyrgyzstan, Pakistan, and the Maldives at the conservative pole. Democratic commitment and anti-elite sentiment place high-trust consolidated democracies and authoritarian states nearer each other than a single-trait reading would, which is why we attach no left--right pole to either. Apart from the health-care item, the Afrobarometer carries no economic left--right attitude item; we verify this against all 300 variable labels of its ninth round.

\paragraph{Coverage.} Of the 155 countries with respondents interviewed since 2013, social liberalism reaches 149, anti-elite sentiment and democratic commitment 154 each, economic left--right 110, and religion in government 93. Economic left--right reaches fewer countries than any dimension except religion in government because the Arab Barometer asks no economic attitude item and the Afrobarometer only one, on the government's responsibility for health care; the items they do ask, on how well the government is narrowing income gaps, rate performance rather than state a position, and we keep them out of the dimension. The six countries without a social liberalism estimate are Algeria, Cuba, Kuwait, Palestine, Qatar, and Yemen.

\subsection*{The religion-in-government dimension}

The religion-in-government dimension measures support for religious authority over law, government and politics, not religiosity. Its items were assembled by a search of every source in the corpus, cross-national and national, first of variable labels in each source's languages and then of the full questionnaires and codebooks, since several qualifying items carry labels with no religious term (Israel's question on conducting public life by Jewish religious tradition is labeled only by its number). Candidates were reviewed one by one against that definition. Items about belief, practice, trust in religious institutions, or attitudes toward other religious groups were excluded, as were four items about religion itself that an earlier version of this dimension had used: whether religion is right when it conflicts with science, whether one's own religion is the only acceptable one, whether religions bring conflict, and whether religious people are intolerant. The items enter the item bank through the same build as every other item: read from each source's raw release, recoded to $[0,1]$ with higher meaning more religious authority, and binned to the coarsest category count among an alignment's members. For the four alignments the bank already held, the build reproduces the bank's respondents exactly before replacing them, which verifies the identifier join for both the cross-national programs and the election studies.

The fit has 22 items, 19 of the dimension's own and three it shares with other dimensions, among them five from the Pew Global Attitudes Survey, on scripture's influence over the laws, scripture against the will of the people, religion as part of national identity, and whether a leader should share the respondent's religion and hold strong religious beliefs; a twentieth of its own, the Danish question on Christmas trees in public spaces, cannot be fitted because the respondents who answer it answer no other religion item. The shared items are the religious-law regime question (World Values Survey and Arab Barometer) and the Asian Barometer's question on consulting religious authorities, both held by democratic commitment, and religious leaders' corruption, held by anti-elite sentiment. Sharing leaves the owning dimension's fit unchanged. Programs are linked by aligned items: the essential-characteristics question on religious authorities and the law (World Values Survey and Asian Barometer), religious leaders' influence on government or votes (Arab Barometer and ISSP), what a country's laws should rest on (Arab Barometer, ISSP, and the Israeli election study), religious rules over national law (ISSP and the Dutch election study), and religion as part of national identity (ISSP, CSES, and the Austrian and Irish studies). Two items were dropped after the first fit because they did not load in their own data whichever way they were coded: the ISSP's question on government interference with religions spreading, and the Dutch question on preventing new mosques, which also reversed the sign of the weakly linked Dutch block. The fit converges on 319,058 respondents answering at least two items, at a marginal reliability of 0.68, with discriminations from 0.37 to 6.16 and one negative ($-0.75$), on religious leaders' corruption, whose direction is set by the anti-elite pole that owns it.

Where two programs cover the same countries, their religion country means correlate at 0.59 to 0.88 and differ in level by 0.04 to 0.15 standard deviations (the six program pairs sharing at least ten countries), which the wave-specific means of the spread model absorb. Because the likelihood spread uses respondents who answer two items, as most World Values Survey respondents do on this dimension, religion spread is measured in all 93 countries the dimension covers. The dimension passes the item-level screen described below in every region where an item can be tested, and leaves its spreads unchanged under it. Its items are too specific to their programs for the regional refit, which needs at least eight items answered in every region.

\subsection*{Respondent estimates and their standard errors}

We score every respondent answering at least two of a dimension's items by expected a posteriori estimation against the fixed item parameters, on a 61-point quadrature grid spanning $[-4,4]$ with a standard normal prior. The same posterior yields a standard error, $\mathrm{se} = \sqrt{E[\theta^2] - E[\theta]^2}$, and that quantity enters every analysis that follows, because every dispersion we report is corrected with it. In all, 2,131,072 people interviewed since 2013 carry an estimate on at least one of the five dimensions.

\subsection*{Polarization, and the corrections it requires}

We report three quantities per country. \emph{Spread} is the standard deviation of individual positions. \emph{Bimodality} is Hartigan's dip statistic \citep{hartigan1985dip}, computed with \texttt{diptest} \citep{maechler2024diptest}; we report the statistic rather than its $p$-value, which rejects unimodality on trivial departures at these sample sizes. \emph{Alignment} is the within-country correlation between a respondent's positions on two dimensions.

We estimate spread directly rather than computing it from individual positions. Within country $c$, a respondent $i$ interviewed by survey wave $s$ has $\theta_i \sim N(\mu_{cs}, \sigma_c^2)$, with the item parameters held at their fitted values, and we maximize the marginal likelihood $\sum_i \log \int \prod_j P(x_{ij} \mid \theta)\, \phi(\theta; \mu_{cs}, \sigma_c)\, d\theta$ over the wave means $\mu_{cs}$ and a common $\log \sigma_c$, on a 201-point grid over $[-5, 5]$. The country's spread is $\sigma_c$, with a standard error from the Hessian; every respondent answering at least one item enters, weighted by the information in their answers. All fits converge; one, social liberalism in Haiti, reaches the lower bound and is dropped. On social liberalism the spread estimates leave out Afrobarometer rounds 6 to 9, whose only items on the dimension are the unwanted-neighbors battery and two weakly discriminating questions: positions resting on them stay close to the prior (median posterior standard error 0.86), and the neighbor items and the rest of the instrument rank countries unrelated to each other (Spearman 0.07). The exclusion removes 734,343 answers and the social-liberalism spreads of Burkina Faso, Burundi, Niger, and Sudan; positions and every other analysis keep these respondents. With the four rounds included, the spreads of the 142 countries estimable both ways correlate with ours at 0.92. The model handles two problems that are not robustness checks, because both change the country ordering.

First, estimation error. Item counts run from two to over sixty, so the precision of individual positions varies widely across countries. Expected a posteriori positions are shrunk toward the prior, most strongly for respondents who answered fewest items, so the familiar correction, subtracting the mean posterior variance from the variance of the point estimates, compounds the shrinkage rather than removing noise. On this corpus it returns a non-positive variance in 7 countries on social liberalism and 30 on democratic commitment, countries whose respondents answer few items (mostly African on social liberalism, mostly Latin American on democratic commitment), and whose spread under the likelihood is ordinary, at about 0.6 to 0.7. The likelihood instead treats error inside the model, and the resulting spread is nearly unrelated to the number of items a country's respondents answered (rank correlation $-0.19$ on social liberalism, $-0.11$ on democratic commitment).

Second, a country's respondents are drawn from more than one survey program, and the programs place the same national population in different locations. We estimate the magnitude by decomposing each country's within-country variance on social liberalism, where it is covered by more than one survey wave, into within- and between-wave parts: the median country places 7.6\% of it between waves, the ninetieth percentile 18.8\%, and the largest single case, Bangladesh, 47.0\%. That these are instrument effects rather than sampling variation is established by holding the pair of waves fixed and varying the country --- the eleventh round of the European Social Survey sits $0.66$ above the seventh wave of the World Values Survey with the same sign in 8 of 8 shared countries, the 2023 ISSP module $1.04$ above the fifth European Values Study wave in 16 of 16, and the 2018 ISSP $0.68$ below the 2023 module in 21 of 23 --- while consecutive waves of a program that repeats its battery differ on average by $0.06$ (Latinobar\'{o}metro), $0.02$ (Afrobarometer) and $0.13$ (European Social Survey), against a median of $0.27$ between waves of different programs. The exception is consistent with that mechanism, on the one case available: the 2018 and 2023 ISSP waves differ by $0.68$, and the ISSP rotates its topical module, hence those two waves pose largely different items. The offset therefore tracks item content rather than the fielding organization, which is the signature of differential item functioning and not of a difference in mode, frame or fieldwork year. The wave-specific means $\mu_{cs}$ absorb these offsets, so that the between-program component of a country's dispersion does not enter $\sigma_c$, while no country mean is moved. We give each survey wave its own mean rather than each program, which is the conservative choice; it removes real change between waves along with the instrument effect, and we cannot separate the two.

The centering is silent about country means by construction. If programs place the same population in different locations, a country's mean is biased by which programs happened to reach it, and adding the country mean back does nothing about that. The design supports an estimate: 122 of the 144 countries with a social-liberalism spread are covered by more than one survey wave, 36 of the 55 waves carrying that dimension span more than one country (it is the wave, not the program family, that fields a given questionnaire), and only 2.0\% of respondents sit in a country covered by a single wave; a two-way model $\theta_i = \alpha_{c(i)} + \gamma_{s(i)} + \varepsilon_i$ is identified. Fitting it, the wave effects span 1.11 standard deviations, from the fifth European Values Study wave at $-0.45$ to the 2021 Canadian election study at $+0.66$, a single-country wave, and the adjusted country ordering correlates with the raw one at $+0.93$ on social liberalism, $+0.96$ on economic left--right, $+0.97$ on anti-elite sentiment and $+0.82$ on democratic commitment; the median country moves 0.19, 0.05, 0.07 and 0.14 standard deviations. The conservative pole is unaffected, with Jordan, Libya, Kyrgyzstan, Pakistan, and Myanmar the five most conservative before and after, and the liberal pole changes in order and in two of its members: Andorra, Iceland, New Zealand, the United States, and Uruguay before, Iceland, Andorra, Canada, New Zealand, and Sweden after (positions here average respondents answering at least three items, so the unadjusted poles differ slightly from those of Appendix~C). We report this as an upper bound and do not adopt it, because $\gamma$ absorbs mode, sampling frame, and fieldwork year along with item functioning, and only the last is an artifact we would want removed. It bounds how far a country's position could move if every between-program difference were measurement; the largest movements belong to single-program countries, whose adjustment is imported entirely from other countries rather than evidenced within their own data.

The standard alternative, the subtraction correction applied to positions centered within survey source, agrees with the likelihood spread where respondents answer many items and poorly where they answer few: their rank correlation across countries is $0.90$ on anti-elite sentiment and $0.80$ on economic left--right, but $0.69$ on social liberalism, $0.41$ on religion in government and $0.28$ on democratic commitment. The most divisive dimension agrees between them in 80 of 147 countries. Because the shrinkage that separates them is a property of the estimator and not of the publics, we report the likelihood spread throughout.

Bimodality carries a limitation neither correction addresses. A latent estimate built from a small number of ordinal items takes a limited number of distinct values, and the dip statistic reads that lattice: across countries it correlates with median item count at $-0.71$ on social liberalism and with the number of distinct values per respondent at $-0.78$, whether or not positions are centered within survey wave, because the problem is discreteness rather than noise or mixture.

We attempted the standard remedy and report that it does not work here. Plausible values --- drawing $\theta_{\mathrm{pv}} \sim N(\hat{\theta}, \mathrm{se}^2)$ ten times per respondent and averaging the dip over draws --- remove the lattice, and the normal approximation to the posterior is adequate, since $\mathrm{Var}(\hat{\theta}) + \overline{\mathrm{se}^2}$ equals 1.005 to 1.050 against a prior variance of 1 on social liberalism, economic left--right, democratic commitment, and anti-elite sentiment. The draws are nonetheless smoothed in proportion to each respondent's standard error, which is largest where the items are fewest, leaving the plausible-value dip correlated with item count at $+0.25$ where the point-estimate dip correlates at $-0.71$. The two versions correlate with each other at $0.21$, and Monte Carlo error is not the cause: the standard deviation of the dip across draws is 0.0007 against a range of 0.0008 to 0.0159. The two statistics bracket the truth from opposite sides without locating it. We therefore report the point-estimate dip, which is what the literature computes, and treat no country ranking on bimodality as interpretable. The three-way disagreement that Appendix~E rests on holds under both versions, and more strongly under plausible values.

Correlations between dimensions are attenuated by the same error, and we disattenuate them by the standard correction before comparison. Countries enter a spread comparison at 400 scored respondents rather than the 100 that suffices for a mean, since a second moment needs more data than a first. Bimodality and the distributions drawn in the figures use only respondents who answered at least three of the dimension's items, since a position resting on two items is mostly prior; spread needs no such floor, because the likelihood weights each respondent by the information in their answers. Because raw spreads on different dimensions are not comparable --- each dimension has its own reliability, item count and scale --- we settle the question of which dimension a country is most divided on with a weighted two-way model, $\log\hat\sigma_{cd} = \alpha_d + \beta_c + \gamma_{cd}$, fitted by weighted least squares. The residual $\gamma_{cd}$ is mostly the country-by-dimension variation the rule exists to find (standard deviation $\tau = 0.24$ in log spread) rather than sampling error (median 0.017), so we weight each cell by $1/(\tau^2 + \mathrm{se}_{cd}^2)$, with $\mathrm{se}_{cd}$ the delta-method standard error of the log spread and $\tau^2$ estimated by the method of moments, as in a random-effects meta-analysis; weighting by sampling precision alone would let the most heavily surveyed countries define the average country. The dimension effects $\alpha_d$ absorb each scale's arbitrary unit and the country effects $\beta_c$ how divided a public is in general; the identifying assumption is that the residuals $\gamma_{cd}$ average to zero within each dimension, so that across countries, adjusting for which countries each dimension covers, no dimension is more divisive than another. A country's sharpest division is the dimension with the largest $\gamma_{cd}$, among those it is measured on, if it has at least two; ties break in the order social liberalism, economic left--right, democratic commitment, anti-elite sentiment, religion in government. Its winner probability is the share of 1,000 draws (seed 20261106) in which the same dimension wins when every log spread is redrawn from its sampling distribution and the model refitted. The rule lets countries measured on few dimensions in, at a cost: a country measured on two dimensions has fewer chances to show a sharp division on one. That cost falls on anti-elite sentiment in particular, because anti-elite spread is larger in countries covered by fewer programs ($\rho = -0.32$ between spread and program count), and the countries measured on few dimensions are mostly single-program African countries. Restricting the comparison to the 85 countries measured on all five dimensions reproduces the modal dimension in Europe, Latin America and the Middle East; too few African countries remain to say anything about Africa.

\subsection*{Party polarization, party sorting, issue salience, and expression}

\emph{Party-system polarization} uses V-Party v2 \citep{lindberg2022vparty,dupont2022global}, which codes parties winning at least 5\% of the vote at each national election to 2019. Each dimension is matched to two V-Party indicators: LGBT social equality and immigration (social liberalism), economic left--right and welfare (economic left--right), the anti-pluralism index and political pluralism (democratic commitment), anti-elitism and the populism index (anti-elite sentiment), and religious principles alone (religion in government). Each indicator is standardized across party-elections from 2010 on, the second of each pair is oriented by the sign of its correlation with the first, and the two are averaged into one position per party. A country's polarization on a dimension is the vote-weighted standard deviation of those positions at its most recent election between 2010 and 2019 with at least two parties carrying both a position and a vote share \citep{dalton2008quantity}; a standard deviation has no direction, so the polarity of each V-Party scale does not matter. The sharpest dimension follows the two-way rule for publics, fitted to log party polarization with party standard deviations floored at 5\% of the dimension's median (a few party systems take nearly identical positions), and is taken over the dimensions on which a country is measured in our data, and chance agreement is computed from the two marginal distributions. Code: \texttt{22\_party\_polarization\_vparty.R}.

\emph{Party alignment} is the share of a dimension's within-country variance explained by party, as $\varepsilon^2 = (SS_B - (k-1)\,SS_W/(n-k))/SS_T$ for $k$ parties and $n$ respondents, which corrects $\eta^2$ for the chance alignment that many small parties produce; parties with fewer than 30 respondents are pooled into one residual category, and a cell needs 200 respondents. We compute it within each country, program, and party measure (vote choice, closeness, or in Latinobar\'{o}metro support for the government or the opposition) and average within country, weighting by respondents, over the cells with at least three party groups, on the source-centered estimates. The cross-national programs and the CSES supply party for 939,212 respondents in those cells, across 144 countries, 166,853 of them through the CSES; the single-country election studies are not included. Party enters as a label rather than a position, so the measure asks whether supporters of different parties hold different views and not which party is to the left.

\emph{Sect.} The comparison of sect with attitudes and with party uses Arab Barometer Waves VII and VIII, which ask Muslim respondents their denomination directly; we use that item rather than the harmonized religion covariate. Iraq and Lebanon are the only countries in either wave with at least 100 Sunni and 100 Shi'a respondents. Sect is coded as Sunni or Shi'a; respondents who are not Muslim, including Lebanon's Christians and Druze, are not asked the item, and Muslims who answer ``just a Muslim'' or name another denomination are excluded from both statistics. This leaves 1,311 Sunni and 2,321 Shi'a respondents in Iraq and 1,268 and 1,320 in Lebanon. The share of a dimension's within-country variance explained by sect is the same $\varepsilon^2$ as for party, on the source-centered estimates, with the two waves pooled within country. Its association with party is Cram\'{e}r's $V$ between sect and the party a respondent feels closest to, dropping parties named by fewer than 10 respondents rather than pooling them, on the 1,263 respondents in Iraq (22 parties) and 1,116 in Lebanon (7) who report both.

\emph{Issue salience} comes from most-important-problem questions in the Afrobarometer (rounds 9 and 10), the Arab Barometer (waves 7 and 8), the Asian Barometer (waves 4 to 6), the ISSP 2020, the AmericasBarometer 2023, Latinobar\'{o}metro 2023 and 2024, nine national election studies (Australia, Canada, Finland, Germany, Iceland, the Netherlands, Switzerland, Taiwan, and the United Kingdom), the autumn 2024 Eurobarometer \citep{data_za8905}, and the 2024 Austrian election study \citep{data_autnes}, 131 countries in all. The American election study's coded open-ended answers are not released, so the United States drops out. We classify each of the programs' answer codes, and the open-ended answers of the Canadian, German, and Finnish studies, into economy, public services, security, corruption, other governance, democracy and rights, social and cultural issues, environment, and other, by ordered keyword rules in five languages; 20\%, 7\%, and 45\% of the open-ended answers in those three studies remain unclassified. Four classes map onto dimensions: economy onto economic left--right, social and cultural issues onto social liberalism, democracy and rights onto democratic commitment, and corruption onto anti-elite sentiment. The \emph{first} measure is the share of first mentions in a class; the \emph{any} measure is the share naming the class among up to three mentions of the same question, and also takes the unranked pick lists of the Eurobarometer and the Austrian study. A closed list that does not offer a class records that class as missing for its source rather than as zero, since nobody can name a problem they are not offered.

\emph{Regime and expression.} Regime measures are V-Dem's liberal democracy index \citep{data_vdem16,data_vdem_mm} and its Regimes of the World classification \citep{luhrmann2018row}, taken from Our World in Data \citep{data_owid_regime} and averaged over 2016--2025 for each country rather than matched to fieldwork year. Nonresponse is counted per respondent and dimension as items answered don't-know or refused against items answered, in the cross-national programs; Latinobar\'{o}metro 2024 stores these codes as Stata extended missing values, which a standard conversion discards, and we recover them from the original release. The nonresponse model is a regression of the don't-know share in each country, program and dimension cell (1,636 cells) on lower liberal democracy, standardized, with program and dimension fixed effects, weights for the number of answers, and errors clustered by country. The expression check uses the questions in the Afrobarometer's ninth round and in Latinobar\'{o}metro that ask whether people must be careful about what they say on politics, or whether expressing opinions publicly could have consequences, and compares, within country, the standard deviation of estimates among respondents who say so with the standard deviation among those who do not.

\subsection*{Constraint, and conditioning on it without circularity}

A country's constraint is the within-country correlation between a respondent's positions on two dimensions, computed only where at least 200 respondents are scored on both. The interest gradient splits a country's respondents into terciles of political interest, ranked within country because the item is asked on different scales by different programs, and requires at least 100 respondents in each tercile.

Reporting the gradient separately for countries that exhibit constraint and countries that do not raises a circularity: a country's overall correlation is a weighted blend of the three tercile correlations being compared, and conditioning on it would therefore condition on the outcome. We assign the group out of sample instead. We split each country's respondents at random into halves, compute the overall correlation on the first and the terciles on the second, and repeat over 25 splits. Under this design the median correlation among countries classified as coherent runs $+0.10$, $+0.19$ and $+0.26$ across the three terciles, and the most interested exceed the least in 79\% of country-splits against 54\% among countries classified as incoherent; the rank correlation between a country's out-of-sample constraint and the size of its gradient is $+0.40$. Figure~\ref{fig:coherence}B displays the full-sample terciles with the group assigned out of sample, which is the same conditioning with more legible lines.

\subsection*{Covariates}

We extract demographics from each source's own release and harmonize them to common schemas: age in years and in five bands, sex, education in three levels, urban or rural residence, income rank within country, religion at two resolutions, party attachment and its strength, and banded internet use. We select variables per source wave rather than per program and verify every recode against the release's value labels rather than inferring it from its codes.

We join respondents to their latent estimates by position within the source file. Because that correspondence is an assumption the whole subgroup analysis rests on, we test it. For each of the 18 cross-national sources whose covariates are joined this way, we take up to six items present in both the raw release and the corpus and correlate the two copies, row by row. On the current corpus all 18 return a median rank correlation of exactly $1.000$, which is what an intact row alignment produces and what a shifted one could not.

Table~\ref{tab:demographics} reports, for each characteristic, how many respondents and countries carry it, what share of the corpus that is, and how many source waves contribute.

Among respondents interviewed since 2013, age, sex, and education are close to universal, reaching 92 to 100\% of respondents in all 155 countries, and coverage falls away fast after that: religion and urban/rural residence 80\%, income rank 42\%, internet use 17\%, party attachment 10\%. None of those subsets is a random sample of the corpus; every subgroup result therefore names the variable it uses and the number of countries that carry it.

Three gaps are more consequential than their percentages suggest. The Afrobarometer asks no household income question at all --- it measures material deprivation through a lived-poverty battery instead, which is a different construct --- and the AmericasBarometer's 2023 merged release carries no usable one either, every candidate variable reaching under a seventh of its respondents; income is therefore missing across Africa and most of Latin America. Latinobar\'{o}metro ships no urban/rural indicator, only a settlement-size band, which we cut at 40,000 and 100,000 inhabitants. Religion, finally, is carried at two resolutions: a ten-category coding that every source supports, and a seventeen-category one for the sources that distinguish more. The finer coding exists because two of its categories carry distinctions the coarse coding hides --- evangelical and Pentecostal Christianity, the most consequential religious division in Latin America and much of Africa, which vanishes when folded into ``Protestant'', and traditional and indigenous religion, which the ten-category coding would fold into ``other''. Every fine category rolls up into exactly one coarse one; results can therefore be reported at either resolution without the two halves of the corpus diverging.

The table's last column counts source waves rather than programs, because variable naming drifts between waves of the same program in ways that make per-program variable selection unreliable. Afrobarometer numbers respondent gender Q100 in Round 9 and Q101 in Round 10, and its religion item moves from Q95 to Q97 --- where Q95 in Round 10 asks who decides how money is used. The ISSP replaces its \texttt{DEGREE} education variable with \texttt{EDULEVEL}, on a different scale, between the 2019 and 2020 waves.

\paragraph{Income.} Income is not a single construct across programs, which measure economic position in at least three incompatible ways. The European Social Survey supplies household income in within-country deciles and the AmericasBarometer in within-country quintiles, direct income measures already expressed on a comparable scale. The World Values Survey asks respondents to place themselves on a ten-step income ladder, which is a perception. Latinobar\'{o}metro and the Arab Barometer ask whether household income covers the household's needs, which is economic strain and can be reported identically by people at very different incomes. The table therefore carries two rows rather than one, an income rank and an economic-strain measure, and we never pool them. Any subgroup result reported by income says which of the two it uses and on how many countries.

Because the two are not interchangeable and neither covers the corpus, the income-based subgroup comparisons are reported separately for each measure in Table~\ref{tab:income}; Appendix~D reports only the direction of the income gap, scoped to the countries the income-rank measure reaches. That version covers up to 102 countries, only six of them Sub-Saharan African, because the Afrobarometer asks no income question; the economic-strain measure is carried in 60 countries, in Latin America, the Arab world and, through the 2014 European Election Study, Europe, and its comparisons reach at most 47. Neither supports a claim about the world, and averaging them would produce a number with only the appearance of global coverage.

\paragraph{Language-model assistance in classifying denominations, and its validation.} The Afrobarometer, Latinobar\'{o}metro and AmericasBarometer record religion in long denomination lists --- 131 distinct labels between them, in four languages --- and sorting those into categories needs knowledge of world religion rather than string matching. Keyword rules cannot classify Ansar Dine as a Malian Islamic association, the Zionist Christian Church as an African Independent Church, or the United Church of Zambia as a Protestant union. We therefore coded these labels twice and independently: once by keyword rules and once by asking Claude Opus 5, which sees the label text and the survey program and nothing else, and in particular not the keyword answer, so that agreement between the two methods constitutes independent evidence.

We validate the procedure before use, on an answer key the model never sees. The ISSP, European Social Survey, and Arab Barometer publish explicit denomination categories, and we make our mappings for those by hand from their value labels; run against 34 of those labels the model agrees on all 34, including every one of the 33 it marks high-confidence. On the 131 open labels the two methods agree on 125, covering 165,460 of the 166,266 respondent-codes at issue. The six disagreements are judgment calls rather than errors --- four turn on whether Latin American churches calling themselves \emph{evang\'{e}lica} are historic Protestant denominations or Evangelical ones --- and cover 806 respondents, under a tenth of a percent of the corpus. We resolve them by hand, and a hand ruling overrides both methods wherever one exists.

The disagreements were informative in the other direction as well, exposing three defects in our own keyword rules, including one regular expression that matched the final letters of any word ending in \emph{-na} and had been discarding 4,649 AmericasBarometer respondents whose stated religion was ``believes in a Supreme Being but belongs to no religion'' as missing data. As with the item-direction coding described above, this is defensible because classifying a denomination is a factual question about what a church is rather than a political judgment, and because we check it against human coding rather than assume it. We consult no model about any respondent, any political position, or the content of any item.

\paragraph{Political characteristics.} Alongside the standard background variables the corpus carries four political ones. \emph{Party attachment} is recorded as whether a respondent feels close to any party, in 94 countries, and where the instrument asks the follow-up, how close --- none, weak, moderate, or strong --- in 110. The European Social Survey and Asian Barometer both ask strength only of respondents who have already named a party; the non-partisans are therefore not missing data, and we fill them from the filter question rather than drop them; read on its own the strength variable would describe the quarter of respondents with a party and silently discard the rest. The ISSP instead records the party the respondent voted for, placed on a left--right family scale, which is vote recall rather than attachment and is kept as a separate column. We band \emph{internet use} to daily, weekly, occasional, or never in 125 countries; it is the corpus's closest measure of exposure to online political information.

Two further variables serve a distinct function. \emph{Left--right self-placement} is a harmonized item that sits in the unassigned group: we recode and orient it like any other item but admit it to no dimension, and it therefore enters no fit. It is therefore a held-out validator, available for 697,758 respondents in 96 countries. Political interest is held out on the same terms, in 120 countries. Because we assign neither to a dimension, neither faces the orientation check that dimension membership brings; we therefore verify their direction separately. Across programs sharing at least five countries with 100 respondents each, country means correlate positively on every pair, 20 for self-placement and 16 for political interest (median $r$ 0.68 and 0.75); between individual survey waves, 4 of 281 pairs are negative, none below $-0.22$.

\begin{table}[htbp]
\centering\small
\caption{\textbf{Respondent characteristics and their coverage.} Coverage of each harmonized characteristic among the 2,169,644 respondents in the joint corpus interviewed since 2013, a slightly larger base than the 2,131,072 respondents the analyses use. ``Countries'' counts those with at least one respondent carrying the characteristic; ``usable'' those with at least 100, the same floor used for country-level estimates; the two coincide for every characteristic here, so that floor removes no country. Coverage is uneven by construction, because the survey programs ask different background questions: the Afrobarometer asks no household income question and the AmericasBarometer's 2023 merged release carries no usable one, leaving income absent for the African and much of the Latin American block. Income rank and economic strain are reported separately because they are different constructs and are never pooled.}
\label{tab:demographics}
\begin{tabular}{llrrrrr}
\hline
Characteristic & Coding & Respondents & \% & Countries & Usable & Waves \\
\hline
Age (years) & continuous & 2,131,079 & 98\% & 155 & 155 & 71 \\
Age group & five bands & 2,143,964 & 99\% & 155 & 155 & 74 \\
Sex/gender & male / female / other & 2,158,902 & 100\% & 155 & 155 & 74 \\
Education & ISCED low / medium / high & 1,985,461 & 92\% & 155 & 155 & 73 \\
Urban/rural & urban / suburban / rural & 1,726,934 & 80\% & 151 & 151 & 58 \\
Religion (coarse) & 10 categories & 1,738,463 & 80\% & 151 & 151 & 64 \\
Religion (detailed) & 17 categories & 1,738,463 & 80\% & 151 & 151 & 64 \\
Income rank & within-country percentile & 911,545 & 42\% & 102 & 102 & 38 \\
Economic strain & within-country percentile & 237,748 & 11\% & 60 & 60 & 12 \\
Internet use & daily / weekly / occasional / never & 377,790 & 17\% & 125 & 125 & 10 \\
Internet at home & binary & 69,192 & 3\% & 25 & 25 & 3 \\
Has a party & binary & 248,243 & 11\% & 94 & 94 & 7 \\
Party attachment & none / weak / moderate / strong & 223,558 & 10\% & 110 & 110 & 10 \\
Party voted for & far left .. far right & 114,507 & 5\% & 38 & 38 & 5 \\
Left--right self-placement & 0--1, held out of every fit & 697,758 & 32\% & 96 & 96 & 28 \\
Political interest & 0--1, held out of every fit & 925,144 & 43\% & 120 & 120 & 27 \\
\hline
\end{tabular}
\end{table}

\subsection*{Validation against left--right self-placement}

Every other diagnostic we report is internal --- reliability, cross-region agreement of item discriminations, agreement between two country estimators --- and all of them ask whether the model is consistent with itself. Correlating the latent estimates against self-placement asks whether they agree with something the model never saw. We take the correlation \emph{within} country, because a pooled correlation would mix ordering people inside a country with ordering the countries themselves, and only the first is an individual-level validation.

The headline correlations are reported in Appendix~B; their variation across countries and sources is itself informative. On social liberalism the correlation runs $+0.51$ in the United States, $+0.40$ in Canada and Australia, $+0.38$ in Switzerland and $+0.37$ in France, and it is \emph{negative} in Slovakia ($-0.13$), Czechia ($-0.12$) and Bulgaria ($-0.07$). By source it is $+0.27$ in the ISSP and $+0.17$ in the European Social Survey against $+0.13$ in the World Values Survey and $+0.02$ to $+0.07$ in Latinobar\'{o}metro. Economic left--right shows the same shape, more weakly: $+0.45$ in Sweden, $+0.30$ in Denmark and Finland, and $+0.24$ in New Zealand, against approximately zero in Taiwan and under $+0.15$ in Nigeria, Indonesia, and Tunisia.

We read this as a fact about the left--right frame rather than a defect in the scale, and the post-communist cases are the clearest evidence for that reading; the sign reversal in Slovakia, Czechia and Bulgaria is what a literature on those party systems would predict, since ``left'' there carries an association with the former regime that inverts its relationship to social liberalism \citep{marks2006party}. Self-placement is a single item whose meaning is known to vary across countries; a latent scale built from issue positions should thus diverge from it where the label has detached from the underlying issues. The regional screen points the same way from the other side: it finds the social-liberalism items the most stable of the five dimensions' across regions, so where the scale and self-placement part, the label rather than the items is the likelier to have moved.

Two cautions temper this validation. Attenuation is severe --- self-placement is one item and the latent estimates carry real error --- so these correlations are lower bounds rather than estimates of true agreement. The correlation also does not rise with the number of items a respondent answered, which reflects the composition of the item-count bands rather than a defect of the estimates; the high-item bands are dominated by the World Values Survey and therefore by countries where the left--right frame is weak, while the middle bands are more European.

\subsection*{Screening for differential item functioning}

The common metric assumes item parameters hold across countries. Fitting item-by-country effects for 155 countries and 1,378 items is an estimation problem in its own right; we therefore screen instead. We refit each dimension separately within five regional blocks (Africa, MENA, Europe, the Americas, Asia) on a single item set chosen once and shared by every block, and correlate the resulting item discriminations across blocks. Holding the item set fixed makes the comparison interpretable, since letting each region keep its own best-covered items would compare different questions.

One correction is necessary before the correlations are interpretable. Because a unidimensional latent direction is identified only up to sign, a regional fit can come back mirror-imaged with every discrimination negated. None of our twenty regional fits did, but we keep the rule. We therefore anchor each to the full-sample fit and reverse it where its discriminations correlate negatively with it; without this step the screen can report large negative correlations that look like catastrophic item functioning and are an artifact of the sign convention. The African block on economic left--right requires comment, as it correlates with the full-sample fit at only $+0.02$, essentially no relationship in either direction. Its agreement with the other regions, $+0.10$ to $+0.51$, should be read as weak rather than as evidence of shared functioning; on this screen, at this resolution, we cannot say how economic left--right items function in that block.

This is a screen rather than a correction; it cannot license the invariance assumption, but it localizes where the assumption strains, which is more useful than a global fit statistic.

\paragraph{An item-level screen.} The discrimination screen compares whole fits; a second screen asks whether each item behaves as a member of its dimension in each block. For every item and block with at least 300 answers we correlate the response with the respondent's estimate on the full-sample parameters, and the item fails in that block if the correlation is below 0.10 or opposite in sign to its pooled correlation. Items failing in at least one block number 12 of 238 on social liberalism (3\% of its answers), 32 of 215 on economic left--right (22\%), 31 of 204 on democratic commitment (9\%) and 21 of 311 on anti-elite sentiment (3\%); no religion item fails anywhere it can be tested. We then re-estimate every country's spread by maximum likelihood on the fixed parameters twice, once dropping each failing item only in the blocks where it fails and once dropping it everywhere, and recompute the most divisive dimension with the two-way rule; the second run does not re-estimate religion in government, whose spreads it carries over unchanged. Spread rankings correlate with ours at 0.997 to 1.000 under the first and 0.995 to 1.000 under the second, and the most divisive dimension is unchanged in 143 of 154 countries under the first (141 of the 146 measured on the same dimensions before and after) and in 122 of 154 under the second (102 of 110). The second changes more because economic left--right loses nearly a quarter of its answers, and with them its spread in 44 countries.

\subsection*{Reproducibility}

All analysis code is in the replication archive. The pipeline runs in order: corpus construction and orientation checks, dimension membership, item response fits on human responses with every item first oriented by language-model personas, respondent and country position estimation, covariate harmonization, and the analyses and figures reported here. The analysis stages run from one script, which lists each stage, what it writes, and whether its output is out of date; a claims map ties every number in this paper to the stage and file that produce it, and a check recomputes the main text's numbers from those files. We release the harmonized item bank and the fitted item parameters alongside the estimates, because the parameters allow a later wave, or a survey we have not seen, to be placed on this scale without refitting.

\paragraph{Why we do not report a formal test of invariance across programs.} The natural test is a multiple-group item response model with the program as the group, comparing a scalar specification in which every item parameter is common against a metric one in which slopes are common and intercepts free. We attempted it and report that it cannot be estimated here. Most items on social liberalism appear in only one program, counting the national election studies as one, and so have no second program in which to take a different intercept; the fully crossed model is unidentified for most of that dimension's items. Restricting to the items that two or more programs carry does not solve it, because freeing intercepts by program while freeing program means requires anchor items common to every group, and no item in this corpus is carried by every program with enough respondents on those items --- the most widely shared, whether immigrants should be allowed in, reaches five. The scalar model does converge on those shared items. The metric model is not identified, and mirt declines it rather than returning parameters. We therefore rely on the three external diagnostics --- the regional discrimination screen, the within- and between-program variance decomposition, and the two-way country-by-program model --- and do not claim any of them is an invariance test.

The archive also carries the diagnostics behind the corrections above as standalone scripts, so that a reader can check them without rerunning the pipeline: the between-instrument decomposition and the alternative spread specifications, the two-way country-by-program model, the plausible-value bimodality comparison, the per-source direction screen, and the out-of-sample split for the interest gradient. We release every country spread under four specifications --- the maximum-likelihood estimate we report, and three computed from individual position estimates: error-corrected only, error-corrected with instrument centering, and computed within each country's single largest program --- so that a reader who disagrees with our choice can adopt another. On social liberalism the three per-respondent orderings correlate with the maximum-likelihood one at $+0.47$, $+0.47$ and $+0.57$, and the most divisive dimension agrees between the reported measure and the instrument-centered one in 80 of 147 countries; we report the first, for the reasons given above.

\clearpage

\setcounter{figure}{0}\setcounter{table}{0}

\renewcommand{\thefigure}{S\arabic{figure}}\renewcommand{\thetable}{S\arabic{table}}

\FloatBarrier
\section*{Appendix A. Related work and contributions}

Prior work has either harmonized surveys or fitted measurement models, and never both at the individual level worldwide. Ex-post harmonization is mature and has been done at larger scale than ours, with the Survey Data Recycling project spanning 23 international projects and 4.4 million respondents in 156 countries \citep{sdr2023,kolczynska2022combining}, but it delivers aligned variables, and alignment is a claim that two questions mean the same thing rather than a measurement model that places two respondents from two instruments on one scale. The literature that does fit measurement models has fitted them to groups: pooled programs yield country-year estimates \citep{caughey2019policy,claassen2019estimating}, and the one multidimensional cross-national latent model we know of is designed so that individual ideal points are unnecessary \citep{berwick2025modgirt}. Where multidimensional estimates span many countries they are fitted within a single program \citep{allison2021religiosity}, whose instrument was designed as one piece, and where common scales span many countries the objects are parties rated by experts rather than mass publics \citep{delacerda2026global}. The result is a literature that recognizes the multidimensionality of politics but measures it, internationally, in one dimension, at the level of the country, or within a single survey program.

The obstacle is that respondents in different programs never face the same questions, which is known in the scaling literature as a bridging problem \citep{bailey2007comparable,shor2010bridge,jessee2016estimate}; it is solved by items shared across instruments, and \citet{kaufman2026bridge} show that a bridge can be constructed even where no shared item exists, from synthetic respondents who answer every questionnaire. We use such respondents more narrowly, to orient items before fitting (Methods). The problem is acute in our setting, because an Afrobarometer respondent and a European Social Survey respondent answer disjoint questionnaires designed with no reference to each other. The one prior item response study spanning the barometer families we use fitted each data set separately and concluded that political trust is not cross-nationally equivalent \citep{vandermeer2019putting}. That warning bears directly on our design and we return to it in the Discussion, but testing it requires the common scale that separate fits cannot provide.

The scale further yields a description of the world's political attitudes at the individual level. Most political disagreement lies within countries rather than between them, at 58 to 91\% of the variance depending on the dimension, and two countries with the same average can differ threefold in how internally divided they are. The demographic characteristics most studied in comparative politics (age, gender, education, urban residence, income, and religion) are more weakly associated with political attitudes than their prominence in the literature suggests, and rarely in the same direction in two countries; the clearest exception is by age, as the gender gap in social liberalism is steadily wider in younger age groups, from $-0.005$ among the over-65s to $-0.051$ among the under-30s --- though a single cross-section cannot tell a generational difference from one that closes as people age. It also lets us test, across ninety-six countries at once, whether ideology coheres within a person, a claim about individuals \citep{converse2006nature} that has been tested mostly in the United States and in European democracies \citep{warncke2025what}. The near-zero median correlation between social and economic positions reported in the main text extends to almost every pair of dimensions, but the median conceals the shape of the distribution: 24 of ninety-six countries exceed $+0.1$ and the United States reaches $+0.54$, and the top of the ranking is led, without being told to, by the publics on which the constraint literature was built, though not only by them. In the coherent countries where political interest is measured, ten of eighty-nine, the correlation rises from $+0.12$ among the least politically interested third to $+0.28$ among the most, as sixty years of theory predicts, while elsewhere the dimensions are unrelated at every level of attention. Both of Converse's claims hold; what does not hold is the generalization from the ten countries in which they can be observed.

Finally, the scale gives polarization a comparable definition. Existing cross-national measures rest on party positions or on a single left--right self-placement item whose meaning varies across countries \citep{king2004enhancing,bauer2017leftright}, whereas a latent position estimated from a common item bank yields a dispersion that is the same quantity in Chile and in Ghana. Polarization has long been described conceptually as several quantities \citep{dimaggio1996have,lelkes2016mass}; we show empirically that spread, bimodality and the alignment of positions across dimensions rank countries nearly independently. And we ask whether the divisions we measure are the ones politics is organized around, by setting each dimension against party choice and against what respondents name as their country's most important problem. The answer is uneven, and it bounds what attitudinal measures, ours included, can say about polarization as a political process. One further use of the scale is developed elsewhere. Because the item parameters are fixed and human-anchored, any actor who answers the items can be scored into the human distribution without disturbing it, and \citet{kaufman2026models} use this property to place 94 language models in the world distribution. We refer to those results in Appendices F and G.

This study makes three contributions. First, we construct, to our knowledge, the first measurement that places individuals from different survey instruments and different countries on a common multidimensional political metric, at this breadth of countries and topics. Second, we measure political attitudes at the level of people rather than countries. Division is a property of a distribution and a country mean has none; individual estimates therefore open questions that previously could not be asked: how far apart a country's young and old are, whether the educated hold different positions from the school-leavers, what share of Brazilians is more socially liberal than a given reference point. A country mean contains no Brazilians; hence no such share can be computed from it. Third, we release the item bank and the fitted item parameters, allowing respondents from any subsequently fielded survey to be placed on the same scale without re-estimation, once their items are harmonized to the bank.

\FloatBarrier
\section*{Appendix B. The joint corpus and data details}

\subsection*{Composition and linkage}

Table~\ref{tab:corpus} gives the composition of the human corpus. Twelve cross-national programs and the national election studies contribute 1,378 distinct harmonized items and 122,620,321 responses from 3,997,205 respondents in 161 countries, fielded between 1981 and 2025; the analyses use the 2,131,072 respondents interviewed since 2013. The national election studies are the largest single block, contributing 766 items, because they are the only source that asks a wide battery of political questions with country coverage outside the barometer regions.

Harmonization proceeds by alignment rather than by exact question matching. An alignment asserts that two or more items measure the same construct, and we rescale its members to $[0,1]$, orient them semantically, and treat them as a single column. We build alignments by hand and check them in two independent ways described in the Methods. We hold items that survive neither check, and items whose construct fits no dimension, in an unassigned group of 399 items that enters no analysis.

\begin{figure}[htbp]
\centering
\includegraphics[width=0.92\linewidth]{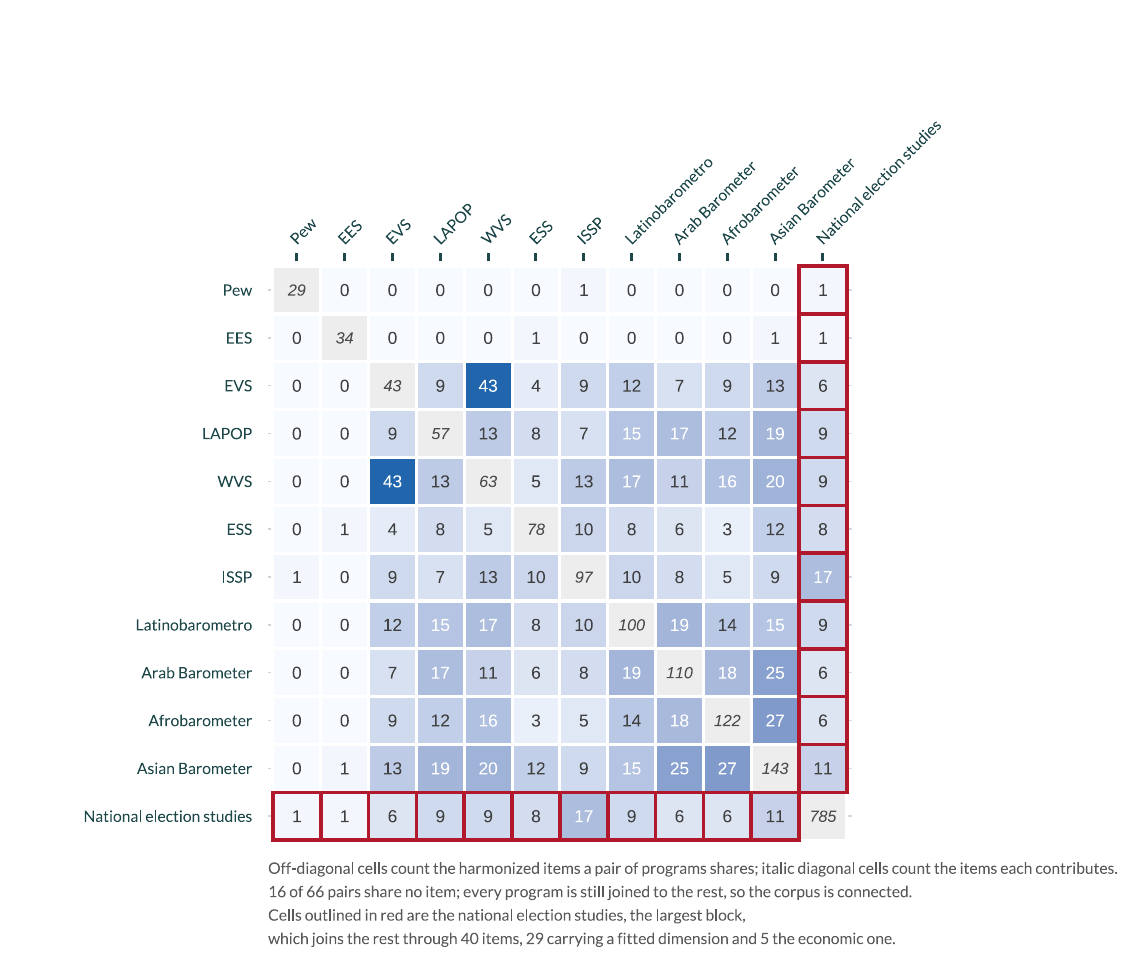}
\caption{\textbf{The linkage structure of the joint corpus.} Off-diagonal cells count the harmonized
items a pair of survey programs shares; italic diagonal cells count the items each contributes.
Sixteen of the 66 pairs share no item, all of them involving the European Election Studies or the Pew survey, but every program is joined to the rest, and the corpus therefore forms a single connected graph. Cells outlined in red are the national election studies, the largest block in the
corpus by items, which join the rest through 40 items in total, of which 32 carry one of the five dimensions and
five the economic dimension.}
\label{fig:linkage}
\end{figure}

Table~\ref{tab:dims} summarizes the five fitted dimensions and the principal quantities reported in the paper. Figure~\ref{fig:linkage} gives the linkage structure, the count of harmonized items each pair of programs shares. Forty items join the national election studies to the cross-national programs. Table~\ref{tab:linkage} names the twenty-nine that carry social liberalism, economic left--right, democratic commitment, or anti-elite sentiment, and marks which dimensions each of them carries; three more carry religion in government, through the Israeli, Dutch, Austrian, and Irish studies and the CSES.

\begin{table}[htbp]
\centering
\caption{The joint human corpus, by source program. ``Items'' counts distinct harmonized item identifiers contributed by that program; an item shared across programs is counted in each. CSES modules are fielded within national election studies. We report them here as a program of their own, because they supply country coverage no single-country study reaches, but we count them inside the national election studies block wherever the linkage structure is analyzed.}
\label{tab:corpus}
\begin{tabular}{lrrrr}
\hline
Source & Respondents & Items & Countries & Responses \\
\hline
ESS & 540,661 & 78 & 39 & 27,999,322 \\
ISSP & 491,042 & 97 & 53 & 7,896,236 \\
Latinobarometro & 489,740 & 100 & 19 & 14,164,371 \\
WVS & 443,017 & 63 & 107 & 16,591,783 \\
LAPOP & 434,978 & 57 & 29 & 12,040,910 \\
NES & 424,376 & 766 & 51 & 6,712,955 \\
Afrobarometer & 303,809 & 122 & 45 & 15,768,482 \\
CSES & 223,209 & 34 & 53 & 2,803,788 \\
EVS & 222,910 & 43 & 48 & 6,205,655 \\
ArabBarometer & 133,531 & 110 & 16 & 3,748,623 \\
PEW & 118,023 & 29 & 36 & 935,663 \\
EES & 109,427 & 34 & 28 & 1,141,663 \\
AsianBarometer & 62,482 & 143 & 16 & 6,610,870 \\
\hline
Total & 3,997,205 & 1,378 & 161 & 122,620,321 \\
\hline
\end{tabular}
\end{table}

\begin{table}[htbp]
\centering
\caption{The five latent dimensions. Fits are graded response models estimated on human data alone, with item parameters then held fixed for scoring. ``Respondents'' is the sample used to estimate item parameters: every respondent, in every wave from 1981 to 2025, answering at least two of that dimension's items. Scoring applies the same rule against the fitted parameters held fixed, to respondents interviewed in 2013 or later only. ``Countries'' counts those with at least one scored respondent since 2013, and ``$\geq$400'' those with a spread estimate, which requires 400 respondents (Methods). Spread is the median across countries of the maximum-likelihood within-country standard deviation, in human standard deviations, with the range across countries in brackets. Economic left--right has fewer countries than any dimension except religion in government because the Arab Barometer asks no economic attitude item and the Afrobarometer only one, on the government's responsibility for health care. One social-liberalism spread, Haiti's, reaches the estimator's lower bound and is excluded, and Burkina Faso, Burundi, Niger, and Sudan have none because the spread estimates leave out Afrobarometer rounds 6 to 9 (Methods).}
\label{tab:dims}
\small
\begin{tabular}{lrrrrrl}
\hline
 & & & & \multicolumn{2}{c}{Countries} & Within-country \\
Dimension & Items & Respondents & Reliab. & any & $\geq$400 & spread \\
\hline
Social lib. & 238 & 2,246,945 & 0.70 & 149 & 144 & 0.71 [0.16--1.25] \\
Dem. commit. & 204 & 2,764,010 & 0.60 & 154 & 154 & 0.86 [0.30--1.42] \\
Anti-elite & 311 & 3,483,383 & 0.83 & 154 & 154 & 0.94 [0.56--1.45] \\
Econ. L--R & 215 & 1,213,617 & 0.74 & 110 & 110 & 0.98 [0.57--2.14] \\
Religion & 22 & 319,058 & 0.68 & 93 & 93 & 0.77 [0.34--1.42] \\
\hline
\end{tabular}
\end{table}

\begin{table}[htbp]
\centering
\small
\caption{\textbf{The 29 items that link the national election studies to the cross-national programs.} Of the 40 items the two blocks share, these 29 carry social liberalism, economic left--right, democratic commitment, or anti-elite sentiment; of the remaining 11, three carry religion in government and eight are unassigned and enter no analysis. Between them the 29 carry 12{,}436{,}296 responses from 3{,}447{,}887 respondents across 158 countries, counting every wave. Several items load on more than one dimension, which is why the per-dimension counts (Soc~9, Econ~5, Dem~7, Anti~13) sum to more than 29. Every economic item comes from a single ISSP module, which is the concrete form of the weak economic linkage discussed in the Discussion.}
\label{tab:linkage}
\begin{tabular}{l cccc}
\hline
Harmonized item & Soc & Econ & Dem & Anti \\
\hline
\texttt{satisfaction\_with\_democracy\_198} & $\bullet$ &  & $\bullet$ & $\bullet$ \\
\texttt{democratic\_authoritarian\_values\_63} &  &  & $\bullet$ & $\bullet$ \\
\texttt{institutional\_trust\_117} &  &  & $\bullet$ & $\bullet$ \\
\texttt{ISSP2018:v13} & $\bullet$ &  &  & $\bullet$ \\
\texttt{Afrobarometer\_R10:Q57A} &  &  &  & $\bullet$ \\
\texttt{democratic\_authoritarian\_values\_66} &  &  & $\bullet$ &  \\
\texttt{ESS10:trstsci} &  &  &  & $\bullet$ \\
\texttt{immigrants\_good\_for\_economy} & $\bullet$ &  &  &  \\
\texttt{immigrants\_increase\_crime} & $\bullet$ &  &  &  \\
\texttt{institutional\_trust\_131} &  &  &  & $\bullet$ \\
\texttt{institutional\_trust\_132} &  &  &  & $\bullet$ \\
\texttt{ISSP2016:v18} &  & $\bullet$ &  &  \\
\texttt{ISSP2016:v19} &  & $\bullet$ &  &  \\
\texttt{ISSP2016:v21} &  & $\bullet$ &  &  \\
\texttt{ISSP2016:v51} &  & $\bullet$ &  &  \\
\texttt{ISSP2016:v7} &  & $\bullet$ &  &  \\
\texttt{ISSP2023:v28} & $\bullet$ &  &  &  \\
\texttt{Latinobarometro\_2023:P33N\_D} & $\bullet$ &  &  &  \\
\texttt{Latinobarometro\_2023:P37CSN\_A} & $\bullet$ &  &  &  \\
\texttt{Latinobarometro\_2023:P37CSN\_B} & $\bullet$ &  &  &  \\
\texttt{level} &  &  &  & $\bullet$ \\
\texttt{one\_party\_rule} &  &  & $\bullet$ &  \\
\texttt{social\_cultural\_policy\_208} & $\bullet$ &  &  &  \\
\texttt{strong\_leader\_over\_parliament} &  &  & $\bullet$ &  \\
\texttt{support\_for\_democracy} &  &  & $\bullet$ &  \\
\texttt{trust\_electoral\_commission} &  &  &  & $\bullet$ \\
\texttt{trust\_legal\_system} &  &  &  & $\bullet$ \\
\texttt{trust\_political\_parties} &  &  &  & $\bullet$ \\
\texttt{WVS7:Q251} &  &  &  & $\bullet$ \\
\hline
\end{tabular}
\end{table}

\subsection*{Religion in government}

Religion in government captures a cleavage that organizes politics across the Middle East and much of Asia: the conflict over religious authority in law and government, of which Islamist--secular polarization is the best-known case. It is measured from the items, across all our sources, that ask about religion's role in law, government, or politics: whether religious authorities interpreting the law is an essential characteristic of democracy (World Values Survey, Asian Barometer), whether candidates should be pre-selected by religious leaders and whether government should consult religious authorities (Asian Barometer), whether clerics should influence government and whether religious leaders should influence votes (Arab Barometer, ISSP), whether laws should rest on sharia or the will of the people, whether religious people should hold office, and whether religion is a private matter (Arab Barometer), whether a country's laws should be based on religion and whether to follow the law or one's religious principles when they conflict (ISSP), whether religion is part of being truly of one's nation (ISSP, CSES, Austria, Ireland), whether Christian values should have a greater role (Norway, Finland, Austria), whether religion is a good guide in politics (Netherlands), whether public life should follow Jewish religious tradition and civil marriage be allowed (Israel), and whether scripture should influence the laws or override the will of the people, whether religion is part of national identity, and whether a leader should share the respondent's religion and hold strong religious beliefs (Pew); it also shares religious leaders' corruption with anti-elite sentiment. The 22-item fit covers 319,058 respondents with a marginal reliability of 0.68 (Table~\ref{tab:dims}), placing Mauritania, Pakistan, and Bangladesh at the pole favoring religious authority and Australia, Germany, and Denmark at the secular one. Its positions, averaged from individual estimates, cover 93 countries, all of which also have a likelihood spread (Methods); the pooled-likelihood estimator places 46 more whose respondents answer a single item.

\subsection*{Orienting items with language-model personas}

Shared items link most of the corpus but not all of it. A survey wave whose items on a dimension are shared with no other wave, directly or through other waves, forms a separate block that the model can fit either way up with the same likelihood: negating the block's discriminations together with its respondents' positions leaves the likelihood unchanged, because $a\theta = (-a)(-\theta)$, the thresholds are untouched, and the standard normal prior is symmetric. Estimation keeps whichever way the block starts, and a block joined to the rest by only one or two items can be held upside down in the same way, at a worse local optimum. The national election studies, each fielding its own questionnaire in one country, are the usual case.

We therefore orient every item before fitting, using language-model personas \citep{kaufman2026bridge}. Seventy-five personas, each a short description of a political outlook, are arranged in five panels of fifteen, one per dimension, running from one pole to the other. Using \texttt{gpt-oss-120b}, each persona answers every item in the bank that can be posed as a standalone question, 1,224 in all, worded without reference to any country and offered with its full list of response options and an instruction to reply with one of them verbatim. Each persona answers each item three times at temperature zero, and we keep the answer that at least two of the three replies give, which 98\% of persona--item pairs have; that answer is mapped onto the item's human scale through the same recode its human answers received. For each item, the rank correlation between a persona's position on its panel and its answer gives the direction in which the item runs, and where that direction is negative we reverse the item's response categories for the fit. Reversing an item's categories reparameterizes the graded response model rather than changing it: it decides which way a separate block enters and where estimation starts, and the data decide the rest. After fitting, the parameters are converted back to the original coding. The personas never enter a fit, and they are not scored or used in any analysis. In a test on social liberalism, fitted to 115,173 respondents including every respondent of the Dutch and German election studies, orienting the items raised the log-likelihood by 906 on the same model and data, and brought both studies' blocks, previously reversed, into agreement with the personas.

Orientation settles a separate block's direction but not its location or scale, which the common standard normal prior sets, as it does for the corpus as a whole. A within-country correlation depends on a block's direction only; a spread taken from a single-country election study also rests on that prior.

Mapping a persona's answer onto the human scale requires the corpus's current coding of each item, and the prompts were written before the last orientation corrections. We therefore test each item against human data alone, without reference to any fit, by correlating every respondent's harmonized answer under the coding the prompts were written against with the same respondent's answer under the current one. The correlation lies beyond $\pm 0.95$ in all but three of 783 survey-wave and item comparisons; for 75 study-specific items since merged into a general one, we compare the earlier item with the merged one. The 136 items whose coding had since been reversed are reversed in the persona answers too, and the persona answers on the 12 items whose status cannot be resolved this way are dropped; eight of those 12 are fitted items. With the coding aligned, persona direction agrees with the sign of the fitted discrimination on 91 to 100\% of strongly loading cross-national items on every dimension.

\subsection*{Agreement with left--right self-placement}

We validate the estimates against left--right self-placement, a harmonized item we assign to no dimension and that therefore enters no fit, available for 697,758 respondents in 96 countries. The two dimensions that claim a left--right reading agree with it and the two that disclaim one do not: within countries, social liberalism correlates with self-placement at $+0.17$ (positive in 68 of 91 countries) and economic left--right at $+0.08$ (positive in 54 of 83), while democratic commitment ($0.00$) and anti-elite sentiment ($-0.07$) sit near zero. Agreement is strongest where the left--right frame is native ($+0.51$ in the United States, $+0.40$ in Canada and Australia) and negative in post-communist countries such as Slovakia ($-0.13$) and Czechia ($-0.12$), where ``left'' retains an association with the former regime that inverts its relationship to social liberalism \citep{marks2006party}. We read this pattern as a property of the left--right label rather than a defect of the scale; source-level correlations and attenuation are discussed in the Methods.

\subsection*{Screening for measurement invariance}

A common metric assumes that item parameters hold across countries, an assumption that is standard in cross-national scaling and known to be imperfect \citep{king2004enhancing,davidov2014measurement,stegmueller2011apples}. We screen for violations by refitting each dimension within five regional blocks and correlating the resulting item discriminations (Methods). The screen is most favorable for social liberalism (median cross-region correlation 0.79), followed by anti-elite sentiment (0.63), economic left--right (0.50) and democratic commitment (0.42); on the last two the disagreement lies between Africa and the other regions, and we accordingly treat country-level quantities on those dimensions as orderings rather than measurements wherever African publics are compared with others, and return to the consequences in the Discussion and Appendix~G.

\FloatBarrier
\section*{Appendix C. Country positions and the distributions underlying them}

\begin{figure}[htbp]
\centering
\includegraphics[width=0.9\linewidth]{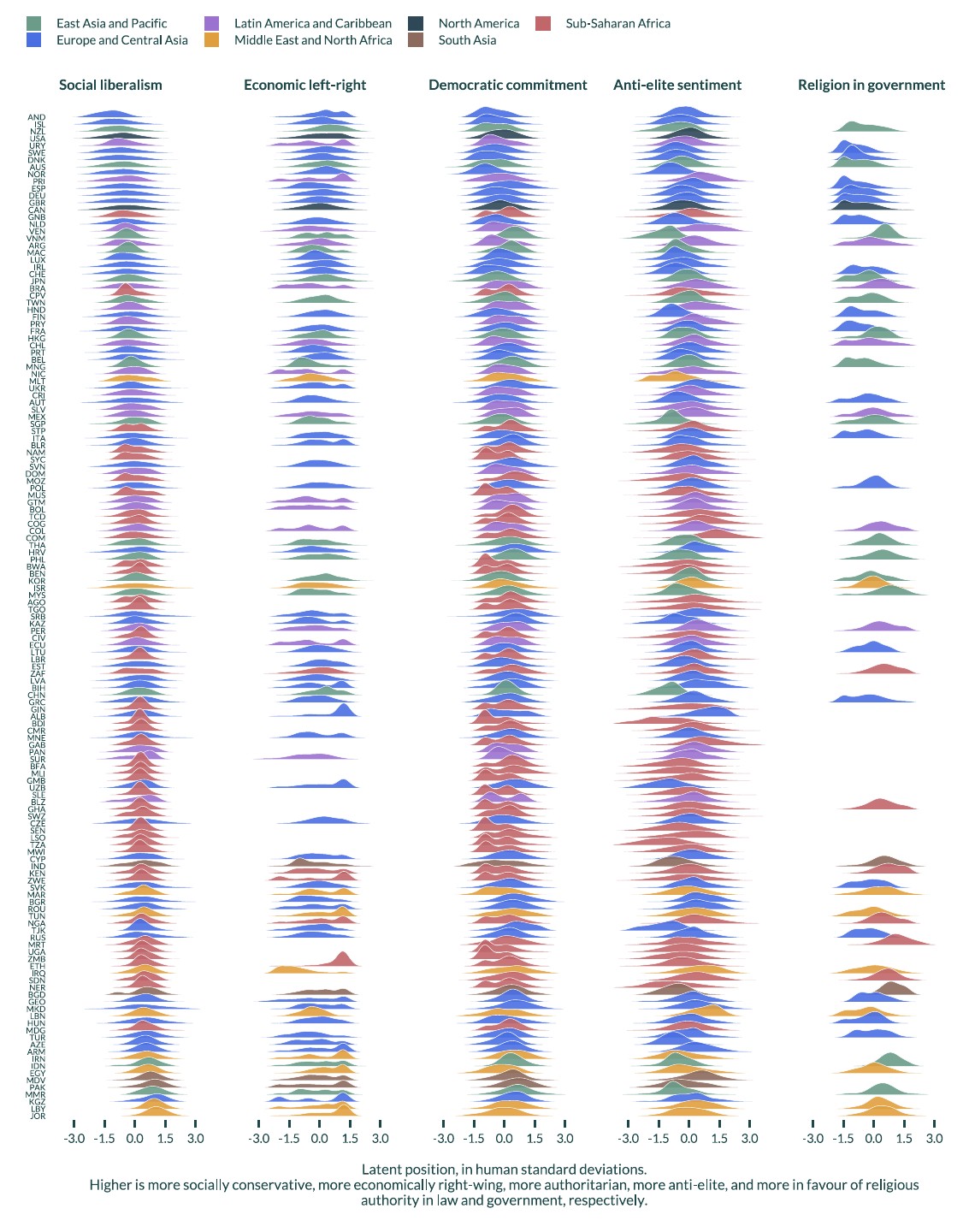}
\caption{\textbf{The political opinion of the world, all 141 countries.} The full version of
Figure~\ref{fig:ridges}, which shows every fourth country for legibility. Each ridge is the
distribution of one country's respondents on one latent dimension, in human standard deviations,
with higher values more conservative, more authoritarian, more anti-elite, or more favorable to
religious authority depending on the dimension. Countries are ordered by mean social liberalism and
that order is held across all five panels; fill marks world region. The figure includes the 141
countries with at least 300 scored respondents on social liberalism, of which economic left--right
covers 98 and religion in government 56. Densities are drawn as
estimated, without correcting for measurement error, which inflates the width of every ridge; the
variance shares quoted in the text are corrected.}
\label{fig:ridgesall}
\end{figure}

Figure~\ref{fig:ridgesall} shows every country we can estimate, on every dimension, drawn as the distribution of its own respondents rather than as a point. Familiar orderings across dimensions are clear --- Andorra, Iceland, New Zealand, Sweden, and the United States at the liberal pole of social liberalism, Jordan, Libya, Kyrgyzstan, Pakistan, and the Maldives at the other --- and the regional structure is visible without being imposed, with Western Europe and Latin America at one end and the Middle East and South Asia at the other. Two regularities extend beyond the country ordering. First, each country is wide relative to the distance between countries, and second, the five dimensions frequently disagree about where a country belongs. The main text examines both.

Country positions are the least novel quantity we measure and the most useful for validation, because they are the only measure that the country-level literature also produces. We compute them two ways. Averaging the individual estimates within a country is the natural estimator but discards anyone who answered a single item, which on economic left--right is most respondents in most sources. Estimating the country mean directly from the pooled likelihood retains those respondents, because the likelihood sums over however many items each person answered. Comparing them provides an internal check. The two estimators correlate at $r = 0.996$ on anti-elite sentiment, $0.99$ on democratic commitment, $0.91$ on social liberalism and on religion in government, and $0.90$ on economic left--right. The comparison is sensitive to item content, which is what makes it useful: it is how we found that the only economic items the Arab Barometer and the Afrobarometer carry rate the government's performance at narrowing income gaps rather than state a position on it, which placed those countries on the economic right under one estimator only. With those items kept out of the dimension, the estimators agree.

A country's position is also exposed to which survey programs reached it, since the programs place the same population in somewhat different locations (Methods). Estimating country and survey-wave effects jointly moves the median country by 0.19 standard deviations on social liberalism and 0.14 on democratic commitment, and leaves the ordering correlated with the raw one at $+0.93$ and $+0.82$ respectively; the poles barely move, with Jordan, Libya, Kyrgyzstan, Pakistan, and Myanmar the most conservative before and after, and Iceland, Andorra, and New Zealand among the five most liberal under either. We report the unadjusted positions throughout, because the program effect absorbs mode, sampling frame, and fieldwork year along with item functioning, and treat the adjustment as a bound on how far a country's place in the ordering could move rather than as a correction to apply.

Most of the world's political disagreement lies within countries rather than between them. Of the variation in social liberalism across the 2.13 million people we estimate, 62\% lies within countries and 38\% between them; on economic left--right the within share is 91\%, on anti-elite sentiment 78\%, on democratic commitment 70\%, and on religion in government 58\%, the dimension on which countries differ most from each other relative to how much their citizens differ among themselves. These shares remove the error in estimating each position and the disagreement between the survey programs covering a country, which is not disagreement among its citizens; with the error removed but the programs left pooled, they run 69, 92, 80, 75, and 59\%. Equivalently, the 149 country averages, laid on the same axis as the people they average, occupy a narrow band inside a distribution several times wider. The country-means literature has only that band to work with. The consequence is not that country means are imprecise but that they answer a different question. Israel and South Korea have the same average position on social liberalism, $+0.08$ against $+0.03$, but their distributions are not remotely alike: Israeli opinion has a spread of 1.07 and South Korean 0.34; Israel is therefore three times as internally divided as a country it is indistinguishable from on the mean.

Two features of measurement can mimic disagreement. A country whose respondents answered few items shows a different observed spread for that reason alone, and a country covered by several survey programs inherits their disagreement with each other as apparent internal division. The spread we report models the first and absorbs the level component of the second (Appendix~E): it is nearly unrelated to the number of items answered (rank correlation $-0.19$ on social liberalism), and Israel and South Korea answered similar numbers of items (six against five on average) from a similar number of survey waves (twelve against eleven); what separates them is therefore disagreement rather than measurement.

\FloatBarrier
\section*{Appendix D. Demographic divisions}

Because a gap between the educated and the uneducated does not exist at the level of a national average, identifying the social divisions opinion runs along requires individual-level estimates. We take six cuts, computed within each country and then compared across countries, since the corpus is not a probability sample of the world and pooling respondents would let its country composition drive the answer (Figure~\ref{fig:cleavages}). The six divisions fall where the comparative literature places them. Age is the largest division on social liberalism, with respondents over 65 sitting 0.25 standard deviations to the conservative side of those under 30, and the religious sit 0.17 to the conservative side of the non-religious. On economic left--right, the top third of earners sits 0.09 to the right of the bottom third. Education is the dividing line on democratic commitment, where the tertiary-educated sit 0.21 toward the democratic pole. On religion in government the religious sit 0.52 toward religious authority relative to the non-religious, the largest median gap in the figure though measurable in only 27 countries, and the tertiary-educated 0.17 toward the secular side, across 48. None of these directions is novel. Because no covariate enters the estimation, the scale recovers these divisions rather than being given them.

Two features of the figure carry more information than the medians. First, the gaps are small. Leaving aside religious affiliation on religion in government, which comes close to measuring the same thing twice, the largest median gap we find anywhere is 0.27 standard deviations, between the top and bottom thirds of earners on religion in government (38 countries), against a median within-country spread of about 0.85; two groups separated by that much, with that much internal variation, still overlap across about six sevenths of each distribution on a normal approximation. Demographic categories sort political opinion weakly, and a respondent's age, education, income, religion, sex, and residence together leave their position largely undetermined. Residence is among the weakest divisions, with a median absolute gap of 0.06 to 0.11 depending on the dimension. That is not at odds with the widening urban--rural divide in how Western democracies vote \citep{rodden2019why,huijsmans2025great}, which describes where parties' votes are concentrated; a modest attitudinal difference, geographically sorted, can produce a large one in winner-take-all systems. The United States is the exception on both counts: its urban--rural gap is the largest in the corpus on social liberalism (0.39, of 131 countries, ahead of Kyrgyzstan's 0.36) and the third largest on economic left--right (0.34, of 88, after Libya's and Georgia's), with city dwellers to the left on both, in the country where the two dimensions are most closely aligned.

Second, the direction of a division is not stable across countries. In the typical division-by-dimension combination, about one country in four has the gap running opposite to the median, and only six of the thirty point the same way in at least nine countries in ten; the direction of the gap, not merely its size, differs between countries. Education on anti-elite sentiment shows this most sharply: the tertiary-educated are more anti-elite than those with only primary schooling in 36 of 39 Sub-Saharan African countries, 14 of 17 in East Asia and the Pacific, and 12 of 15 in the Middle East and North Africa, but less anti-elite in 33 of 44 European and Central Asian countries and in both the United States and Canada. Even the age gap on social liberalism, the most consistent of them, reverses in six of 120 countries. The six exceptions are age, education, and income on social liberalism; education on democratic commitment; and, on religion in government, religious affiliation (27 countries) and income (38 countries, in every one of which higher earners are the more secular). Income on economic left--right does not come close, with 71.9\% of the 96 countries with a measurable income gap on that dimension putting higher earners to the right. Income comes closest to a universal sign, on social liberalism and religion in government --- though it reaches at most 102 countries rather than 155, only six of them Sub-Saharan African (five on economic left--right), and the comparisons themselves are in Table~\ref{tab:income}.

By median absolute gap, sex is the smallest of the six divisions on three of the five dimensions; on economic left--right and religion in government residence is smaller still. Women sit to the liberal side of men on social liberalism (median gap $-0.03$; women the more liberal in 71\% of 141 countries) and to the left on economic left--right ($-0.09$; 85\% of 99). They are slightly less committed to democracy ($+0.04$; on the authoritarian side of men in 78\% of 154 countries) and more favorable to religious authority in government ($+0.06$; 82\% of 61), and they do not differ from men on anti-elite sentiment ($0.00$; women the more anti-elite in 51\% of 154). No median sex gap reaches half the age gap on social liberalism.

\begin{figure}[htbp]
\centering
\includegraphics[width=\linewidth]{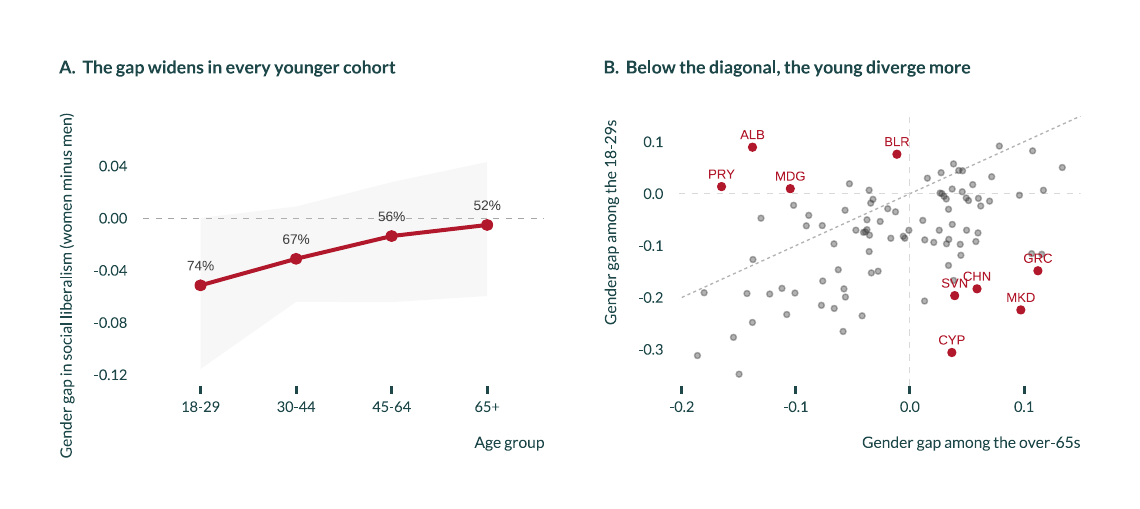}
\caption{\textbf{The gender gap in social liberalism is concentrated among the young.} \textbf{A}, the median gender gap across countries within each age band, with the interquartile range across countries shaded, annotated with the share of countries in which women are the more liberal. A country enters a band only when it has at least 100 women and 100 men in it, which is why the over-65 band covers 96 countries, the 18--29 band 137, and the other two 139 and 138. \textbf{B}, each country's gap among 18--29 year olds against its gap among the over-65s, for the 96 countries with both. Points below the diagonal are countries where the young diverge more than the old; labeled points are the five largest divergences and the four clearest reversals.}
\label{fig:gendercohort}
\end{figure}

\paragraph{The gender gap is concentrated among the young.} One division warrants closer examination. The claim that young men and young women are pulling apart politically while older cohorts are not has circulated widely, but has been tested almost entirely in rich democracies \citep{nennstiel2025gender,milosav2026youth}. Splitting each country by cohort, the gender gap in social liberalism widens monotonically as the cohort gets younger (Figure~\ref{fig:gendercohort}A). Among the over-65s the gap is smallest, at a median of $-0.005$, with women the more liberal in 52\% of countries. Among 45--64 year olds the gap is $-0.014$ and holds in 56\% of countries, among 30--44 year olds $-0.031$ and 67\%, and among 18--29 year olds it is $-0.051$ and holds in 74\% of 137 countries.

The gradient is common but far from universal, and the exceptions are informative (Figure~\ref{fig:gendercohort}B). Thirty-five of the 137 countries in the youngest band run the other way, with young women the more conservative of the two --- among them Bangladesh, Kyrgyzstan, Colombia, Thailand, and Ireland. The gap among the young is widest in Finland, Sweden, Cyprus, Canada, and Bosnia and Herzegovina, and the young diverge most from the old in Cyprus, North Macedonia, Greece, China, and Slovenia.

These results share one limit. Because our data are a single cross-section, a cohort difference and a life-cycle difference are not separable; we cannot tell whether today's young women differ from today's young men in a way that will persist, or whether gender gaps close as people age. The cross-section shows the pattern the topical claim asserts but cannot adjudicate its mechanism; distinguishing the two requires repeated waves scored on a common scale, which the harmonization developed here makes possible as new survey rounds arrive.

\begin{table}[htbp]
\centering
\caption{\textbf{Economic position and political attitudes, by measure.} The corpus measures
economic position two ways and they are not interchangeable. Income rank within country is
available in 102 countries; economic strain, a lived-deprivation battery, in 60. No respondent
carries both, so the two are reported separately and never averaged. Gaps are the difference in
mean latent position between the top and bottom third of each measure, in human standard
deviations, for countries with at least 100 respondents in each group; the bracket gives the
tenth and ninetieth percentiles across countries. The economic-strain comparisons reach at most 47
countries, and 15 on economic left--right. Positive means the top third
by income, or the most strained, sits higher on the dimension.}
\label{tab:income}
\small
\begin{tabular}{llrrl}
\hline
Measure & Dimension & Countries & Median gap & 10th--90th \\
\hline
Income, top vs bottom third & Social liberalism & 100 & $-0.164$ & $[-0.34, -0.02]$ \\
 & Economic left--right & 96 & $+0.091$ & $[-0.09, +0.29]$ \\
 & Democratic commitment & 102 & $-0.113$ & $[-0.28, +0.08]$ \\
 & Anti-elite sentiment & 100 & $-0.131$ & $[-0.32, +0.18]$ \\
 & Religion in government & 38 & $-0.268$ & $[-0.47, -0.05]$ \\
Economic strain, most vs least & Social liberalism & 33 & $+0.090$ & $[-0.03, +0.15]$ \\
 & Economic left--right & 15 & $-0.071$ & $[-0.25, +0.06]$ \\
 & Democratic commitment & 32 & $+0.091$ & $[-0.05, +0.18]$ \\
 & Anti-elite sentiment & 47 & $+0.241$ & $[+0.06, +0.53]$ \\
 & Religion in government & 13 & $+0.041$ & $[-0.06, +0.17]$ \\
\hline
\end{tabular}
\end{table}

\FloatBarrier
\section*{Appendix E. The polarization measures and their corrections}

Polarization is the clearest case of a question the country-level literature cannot answer. It is a statement about the shape of a distribution, and the measures in wide comparative use capture something else --- the distance between party positions, which is a property of elites, or the dispersion of a single left--right self-placement item, whose scale points are known not to mean the same thing in different countries \citep{king2004enhancing,bauer2017leftright}. A latent position estimated from a common item bank gives a dispersion that is the same quantity in every country, which a comparative polarization measure requires.

\begin{figure}[htbp]
\centering
\includegraphics[width=\linewidth]{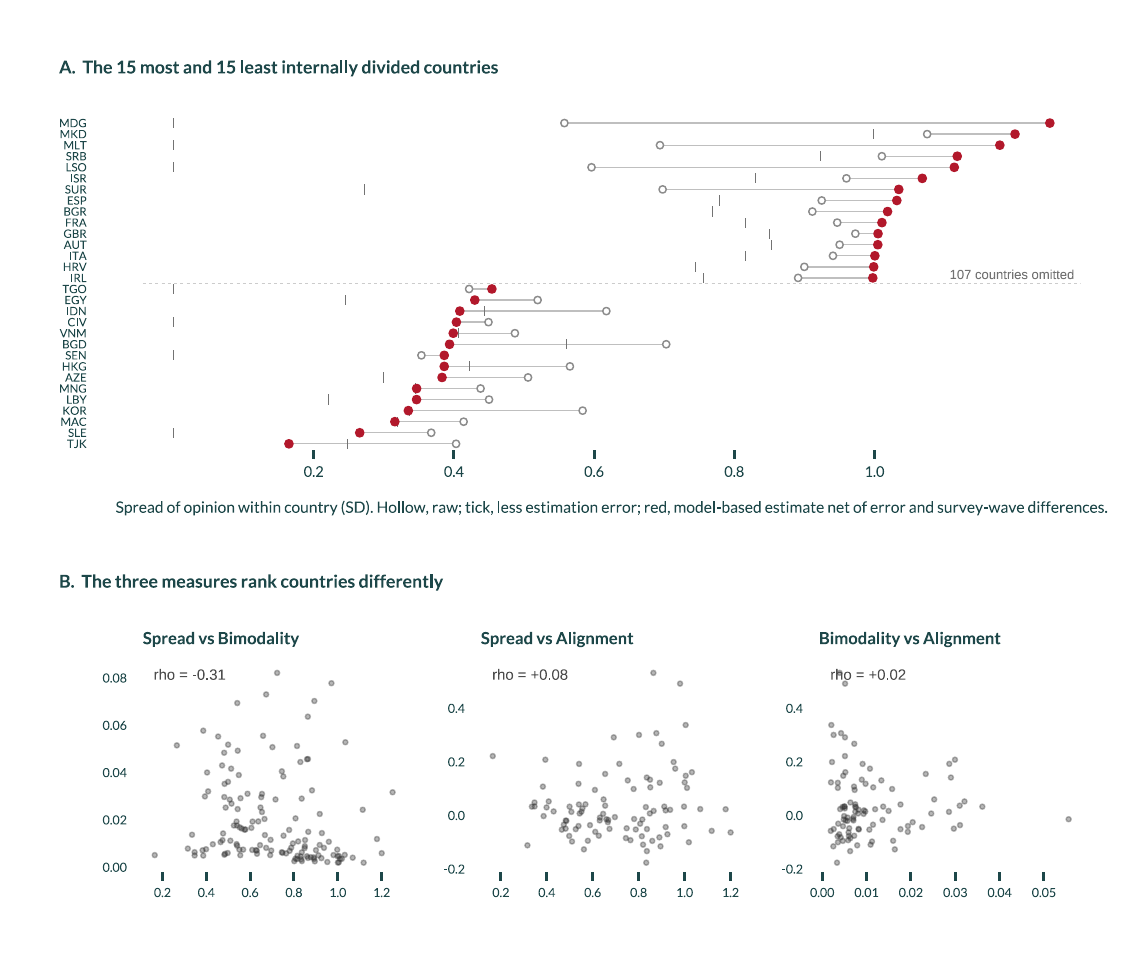}
\caption{\textbf{Polarization is three things, and they rank countries differently.} \textbf{A}, the fifteen most and fifteen least internally divided countries on social liberalism, of 137 with at least 400 respondents scored on three or more items. Hollow points give the raw within-country standard deviation of individual position estimates, the tick the value after subtracting mean posterior error variance, and the red point the maximum-likelihood spread, which models measurement error and gives each survey wave its own mean (Methods). \textbf{B}, the three measures plotted against one another, one point per country, annotated with Spearman rank correlations. Bimodality is Hartigan's dip statistic, computed on positions centered within survey wave; alignment is the within-country social--economic correlation of Figure~\ref{fig:coherence}A, which is why it covers 96 countries rather than 137.}
\label{fig:polarization}
\end{figure}

Spread, bimodality and alignment do not agree (Figure~\ref{fig:polarization}B). Across 137 countries the rank correlation between spread and bimodality is $-0.31$; across the 96 countries where alignment can be computed, the correlation between bimodality and alignment is $+0.02$ and between spread and alignment $+0.08$. A country that looks polarized on one of these measures is not, in general, the country that looks polarized on another. Work that measures one of them and calls it polarization measures one of three distinct quantities rather than approximating a single underlying construct, as conceptual work on polarization has long argued \citep{dimaggio1996have,lelkes2016mass}.

On spread, the most internally divided publics on social liberalism are Madagascar, North Macedonia, Malta, Serbia, Lesotho, and Israel, and the least divided are Tajikistan, Sierra Leone, Macau, South Korea, Libya, and Mongolia (Figure~\ref{fig:polarization}A). The least divided end mixes East and Central Asian publics with two African and Arab ones; the most divided end mixes contested European democracies with smaller states, several of them covered by few programs, and we read neither end as a finding in itself.

Two problems separate a country's spread from the dispersion of its respondents' estimated positions, and both change the ordering.

Measurement error is the first. A respondent's estimated position carries a standard error that shrinks with the number of items they answered, and item counts vary enormously across our sources: a median respondent answers five to eleven items on a dimension, but the range runs from two to over sixty. The usual correction subtracts the mean posterior error variance from the variance of the estimates. With expected a posteriori estimates it fails, because those estimates are not noisy around the truth but shrunk toward the prior, most strongly where items are fewest, so subtracting posterior variance compounds the shrinkage. On this corpus the subtraction leaves a non-positive variance in 7 countries on social liberalism and 30 on democratic commitment, mostly African publics on the first and Latin American ones on the second, answering few items. We instead estimate each country's spread by maximum likelihood from the answers themselves, with the item parameters fixed (Methods), which treats error inside the model. The resulting spreads are nearly unrelated to the number of items a country's respondents answered (rank correlation $-0.19$ on social liberalism), and every country with 400 respondents receives one, except Haiti on social liberalism, whose estimate reaches the lower bound.

A country's respondents are not one sample, and the samples disagree. Most countries here are covered by more than one survey program, and the programs do not place the same national population in the same location. Gaps between waves of different programs run to 1.04 standard deviations, with a median of 0.27, and hold their sign in at least nine shared countries in ten for two fifths of the pairs, while consecutive waves of a program that repeats its battery differ on average by 0.02 to 0.13 (Methods). The one exception is consistent with the mechanism, though it rests on a single case. The ISSP rotates its topical module, hence its 2018 and 2023 waves pose largely different items and differ by 0.68. The offset tracks the items a wave carries rather than the organization fielding it, which is what differential item functioning predicts and what a difference in sampling frame or fieldwork year would not. Pooling programs within a country converts these offsets into apparent disagreement among its citizens. The spread model therefore gives each survey wave in a country its own mean and estimates a single spread around those means, which leaves country positions untouched and removes the between-instrument component of their spread. Figure~\ref{fig:polarization}A plots the raw, the error-subtracted and the model-based value for the most and least divided countries, so that the size of each step is visible.

\emph{Do spread profiles replicate across instruments?} Wave-specific means remove level differences between programs; they cannot remove differences in what the programs' items discriminate. We therefore ask directly whether a country's profile of spreads --- its standing on each dimension relative to the other countries the same wave surveys --- is reproduced by a second program. For each wave we standardize log spread within dimension across the countries it covers (at least five), so that a wave's general tendency to produce wide or narrow spreads cancels, and correlate a country's profiles from two waves over the dimensions both measure (at least three), averaging within country. Two waves of the same program agree at a mean correlation of 0.49 (95\% country-bootstrap interval 0.37 to 0.61, 74 countries) and name the same most divisive dimension in 63\% of countries, against 33\% by chance. Two different programs agree at 0.17 (0.05 to 0.28, 75 countries) and name the same dimension in 41\%. Agreement varies by program pair, from 0.60 between the Asian Barometer and the World Values Survey and 0.31 to 0.35 among the Latin American programs to about zero between the European Social Survey and the World Values Survey, the European Election Studies or the CSES. The same test applied to party alignment gives 0.30 across programs against 0.37 within one (Appendix~H). A country's spread profile is therefore partly a property of the instrument that measured it, and we treat this as a limit on the main text's account of what divides each country rather than as a result. Code: \texttt{24\_profile\_replication.py}.

\emph{Which dimension divides a country most?} Answering requires fixing each dimension's arbitrary unit, and different ways of fixing it embody different assumptions. The main text uses a two-way model of log spread with dimension and country effects (Methods), which assumes that across countries, adjusting for which countries each dimension covers, no dimension is more divisive than another. We compare it with two alternatives. Ranking countries within each dimension, by percentile, assumes the same thing without the adjustment, so each dimension is judged against whichever countries happen to measure it. Comparing spreads in raw units assumes that the world's total variance, within and between countries, is the same on every dimension. The dimension effects of the two-way model are $-0.21$ for social liberalism, $-0.08$ for religion in government, $-0.01$ for democratic commitment, $+0.08$ for anti-elite sentiment and $+0.22$ for economic left--right (log spread, centered), each within $0.03$ of that dimension's raw mean log spread, so adjusting for which countries each dimension covers changes little; what separates the two-way model from percentile ranking is mainly that it uses the size of a country's departure from each dimension's typical spread rather than only its rank. Across the 154 countries the two-way model names social liberalism the sharpest division in 49, anti-elite sentiment in 35, economic left--right in 33, democratic commitment in 27, and religion in government in 10. Percentile ranking names social liberalism in 42, anti-elite sentiment in 35, economic left--right in 31, democratic commitment in 25, and religion in government in 21. Raw units name economic left--right in 63 and social liberalism in only 22, which reflects how each scale's unit is set rather than anything about the countries. The two-way winner agrees with the percentile winner in 86\% of countries and with the raw winner in 69\%, and 98 countries have the same winner under all three; among them are 17 of the 31 European countries that divide most over social liberalism, 14 of the 16 Latin American countries that divide most over economic left--right and all 24 African countries that divide most over anti-elite sentiment. Within the two-way model, redrawing every spread from its standard error leaves the winner unchanged in at least 95\% of 1,000 draws for 132 countries and in at least 80\% for 143. A fourth way of fixing the units, measuring spread in the response metric of each dimension's items, is the only one that gives an absolute comparison, but it ranks dimensions by how sharply their questions discriminate (median discrimination 1.85 on religion in government against 0.61 on democratic commitment) and so inherits the instrument dependence described above; we do not use it. The country-level results are in \texttt{paper\_which\_dimension\_twoway.csv}, written by \texttt{25\_which\_dimension\_twoway.R}.

Two residual points. Spread remains associated with the number of programs covering a country, positively on social liberalism ($\rho = +0.47$) and negatively on anti-elite sentiment ($-0.32$) and economic left--right ($-0.56$); countries covered by several instruments are also richer, more surveyed and more politically contested, and we cannot separate those differences from an instrument artifact. And the model assumes positions are normally distributed within each country around its wave means, which a strongly two-peaked public would violate; the dip statistic, below, cannot test that assumption on this corpus.

Bimodality carries a limitation neither correction addresses, and we were unable to remove it. A latent estimate built from a small number of ordinal items takes a limited number of distinct values, and the dip statistic reads that lattice along with any real two-peakedness: across countries it correlates with the median number of items answered at $-0.71$ on social liberalism and with the coarseness of the lattice at $-0.78$, whether or not positions are centered within survey wave, because the problem is discreteness rather than noise or mixture. The standard remedy is to compute the statistic on draws from each respondent's posterior rather than on the point estimate. Doing so removes the lattice and introduces its mirror image: the draws are smoothed by an amount proportional to each respondent's standard error, which is largest exactly where the items are fewest, and the resulting dip correlates with item count at $+0.25$ rather than $-0.71$. The two versions of the statistic correlate with each other at $0.21$. This corpus therefore cannot support a country ranking on bimodality.

This appendix's central claim survives the ambiguity, because it is a claim about disagreement rather than about any country's rank. The three measures disagree under either version of the dip, and rather more under the plausible-value version: spread with bimodality runs $-0.31$ on the point estimates and $-0.49$ on the draws, bimodality with alignment $+0.02$ and $-0.15$. Whichever dip a reader prefers, it does not rank countries the way spread or alignment does. We report the point-estimate version in Figure~\ref{fig:polarization} because it is the statistic the literature computes, and treat the country ordering on that panel as uninterpretable.

The invariance assumption bears on the polarization measures differently from the other results. A comparison of spreads is a comparison of second moments, which is more demanding than a comparison of means, since item functioning that shifts a country's location is harmless to a spread measure while functioning that changes an item's discrimination changes the spread directly. The regional screen described in Methods is a screen on discriminations, and it is strongest on social liberalism, the dimension Figure~\ref{fig:polarization} uses for illustration (median cross-region correlation 0.79), and weakest on democratic commitment and economic left--right. The central claim does not depend on that choice. Recomputed on each of the other four dimensions, taking alignment as the within-country correlation with social liberalism, the three measures disagree there too: spread with bimodality, spread with alignment, and bimodality with alignment run $-0.25$, $-0.23$ and $-0.03$ on anti-elite sentiment, $+0.18$, $+0.12$ and $-0.43$ on democratic commitment, $+0.44$, $-0.11$ and $-0.18$ on economic left--right, and $+0.04$, $+0.36$ and $-0.40$ on religion in government (61 countries for the first, 29 for the other two). No pairing on any dimension exceeds 0.44 in absolute value on the point-estimate dip. That polarization is several separable things is therefore a finding about the concept rather than an artifact of the dimension we chose to illustrate it with.

The country ranking is the part that should be read cautiously. The regional screen gives least cause for concern on social liberalism and most on democratic commitment and economic left--right, whose discriminations agree least between Africa and the other regions; and the dependence of spread profiles on the instrument, above, applies to every dimension. We would therefore treat any country ordering on spread as indicative and the disagreement among measures as the result.

\FloatBarrier
\section*{Appendix F. Ideological constraint}

Ideological constraint is one of the oldest claims in political behavior and one of the least tested outside the West. \citet{converse2006nature} argued two things: that most people do not organize their opinions along a common line, and that those who do are the ones who follow politics. Sixty years on, both claims have been tested mostly where the data are, in the United States \citep{ansolabehere2006purple,ansolabehere2008strength} and in European democracies \citep{warncke2025what}. Because constraint is a within-person relationship between positions, it cannot be computed from country means, and the cross-national record has therefore left it largely unexamined. On a common scale the pair becomes a single question with a geography: where do the dimensions move together within individuals, and among whom?

\subsection*{Correlations are near zero in most countries and substantial in a few}

The estimand is the correlation between one person's positions on two dimensions, computed within a country so that the statistic is comparable across countries. Figure~\ref{fig:coherence}A gives all ten pairs, one point per country. The median is near zero and the distribution is not. The median within-country correlation between social liberalism and economic left--right --- the canonical test of ideological constraint, and the relationship a left--right axis is supposed to describe --- is $+0.01$ across 96 countries, and it is positive in 52\% of them, a share no different from chance. This agrees with the one earlier test at comparable breadth, which found across ninety-nine nations that cultural and economic attitudes are rarely organized along a right--left line and are more often negatively correlated \citep{malka2019cultural}; our 48\% negative share leans neither way. For most of the world a person's social liberalism carries essentially no information about their economic position. The other nine pairs of dimensions behave similarly; their medians fall between $-0.11$ and $+0.18$, the largest being religion in government with democratic commitment. The least stable is between democratic commitment and anti-elite sentiment, whose median of $+0.09$ conceals a range from $-0.18$ to $+0.78$ across the middle 80\% of countries. Two dimensions whose correlation varies this widely do not combine into a stable composite, and we attach no left--right pole to either.

Religion in government is the one dimension that moves with others, and the way it does answers whether it is a new axis or social liberalism under another name. In the countries where it can be paired with them, support for religious authority goes with the authoritarian pole of democratic commitment (median $+0.18$, positive in 93\% of 42 countries), runs mildly against anti-elite sentiment ($-0.10$, positive in 13\% of 38), and is unrelated to social liberalism ($0.00$, 29) and to economic left--right ($+0.03$, 14). It is therefore not social liberalism under another name. A correlation of $0.18$ leaves over 96\% of each dimension's within-country variance unexplained by the other, so religion in government and democratic commitment are related positions rather than one position measured twice. Religion pairs with other dimensions in fewer countries because fewer surveys ask its items alongside theirs.

The social--economic distribution has a substantial right tail, and the countries in it are not a random draw. Ranked from the top, the correlation is $+0.54$ in the United States, $+0.48$ in Canada, $+0.37$ in the Netherlands, $+0.29$ in Pakistan, $+0.27$ in Switzerland and Australia, and $+0.25$ in Austria. Twenty-four of the 96 countries clear $+0.1$ and nine clear $+0.2$. That ranking is partly a validation: its three highest, and 14 of the 24, are established democracies of the kind on which the constraint literature was built, recovered by a scale fitted without reference to any of them and never told which countries were expected to be coherent. But the list is not only those publics: Pakistan, Tajikistan, Bangladesh, Tunisia, and Malaysia all clear $+0.19$. A measure returning zero everywhere would be evidence of attenuation; this one detects constraint where six decades of single-country work locates it, and in a few places that work has not examined.

The left tail is smaller and makes the same point in reverse. Seven countries fall below $-0.10$: Hungary, the most negative in the corpus at $-0.16$ on 4,015 respondents holding both estimates, then Estonia, Poland, Thailand, Bulgaria, Macau, and North Macedonia. Five of the seven are post-communist European publics, and the negative sign is the alignment the European literature on the radical right would predict, in which socially conservative voters sit to the economic left. What varies across countries is therefore not only whether the dimensions cohere but in which direction.

Three explanations for a spurious null can be ruled out. Measurement error does not produce it. Error in single survey items is what once made mass opinion look incoherent, and scaling many items per respondent is the standard remedy \citep{ansolabehere2008strength}, which our estimates apply; the residual error works against a null in any case, since these are correlations between two estimated quantities, each carrying error, and the observed values therefore understate the truth, but disattenuating with the marginal reliabilities raises them by about two fifths, which leaves a median of zero at zero. The model does not impose it: we fit the five dimensions separately rather than jointly, nothing in the estimation constrains them to be orthogonal, and they were free to come back correlated but did not. Pooling survey programs within a country does not produce it either: recomputing every correlation after removing between-program differences (Methods) moves the median to $+0.01$ and moves the most coherent countries in both directions --- Canada rises to $+0.49$, Switzerland to $+0.31$ and Austria to $+0.34$, while the United States falls slightly to $+0.53$ and the Netherlands to $+0.30$ --- while some of the less coherent European publics, Denmark among them, fall.

\subsection*{Coherence is concentrated among the attentive}

Converse's second claim is that constraint is concentrated among the informed and the politically involved. Pooled across all 89 countries with a political-interest measure, the gradient the theory predicts is absent: the median correlation runs $-0.015$ among the least interested, $+0.007$ in the middle and $-0.013$ among the most --- no rise at all --- and 50 of the 89 put the most interested above the least, against the 45 that chance would give.

That flat line is an artifact of aggregation. In the 79 countries without social--economic constraint the dimensions are essentially unrelated at every level of attention, so there is no gradient to detect, and averaging them with the countries that have one conceals it. Splitting countries by whether they exhibit any social--economic constraint separates the two cases completely (Figure~\ref{fig:coherence}B). Among the ten that do, the median correlation runs $+0.12$ among the least interested, $+0.19$ in the middle and $+0.28$ among the most, and eight of the ten place the most interested above the least. Among the remainder it runs $-0.03$, $-0.01$ and $-0.03$, and the most interested exceed the least in 53\% of them. The United States, where the claim originated, runs $+0.01$, $+0.07$ and $+0.31$, a steep gradient from a low base, though its split-half constraint ($+0.09$) falls short of the $+0.15$ threshold.

Because a country's overall correlation is a weighted blend of the three tercile correlations being compared, conditioning on it would condition on the outcome. We therefore assign the group from a random half of each country's respondents and compute the terciles on the other half, repeated over twenty-five splits. The gradient survives: among countries classified as coherent out of sample the medians run $+0.10$, $+0.19$ and $+0.26$, the most interested exceed the least in 79\% of country-splits against 54\% elsewhere, and the rank correlation between out-of-sample constraint and the size of a country's gradient is $+0.40$.

Two artifacts that could produce a flat line where a real gradient exists are also absent. Politically interested respondents might simply answer more items, making their estimates less noisy and their correlations less attenuated --- but the median item count is 12 in all three terciles on social liberalism and 3 in all three on economic left--right. The mean posterior standard error on social liberalism is 0.41 among the least interested and 0.42 among the most, so the two terciles the gradient compares are measured equally well; the middle tercile is measured somewhat better (0.35), which would raise its correlation relative to the others rather than produce the monotone rise we observe. Disattenuating each tercile separately raises all three correlations by a similar factor (to $+0.20$, $+0.29$ and $+0.43$ in the coherent group) and leaves the gradient and the counts unchanged.

\emph{Does attention build constraint along a country's own divisions?} If the gradient reflected a general mechanism, attentive citizens learning whatever bundle their politics offers, it should also appear between the two dimensions a country is most divided on, which need not be social and economic. It does not. For each of 71 countries we take the two dimensions with the largest excess spread under the two-way rule and correlate respondents' positions on them within each interest tercile. Because the sign of a correlation means nothing across dimension pairs, each country's correlation is oriented by its sign in a random half of its respondents and the terciles are measured on the other half, over 25 splits. The two dimensions are bundled, at a median of $+0.16$ against $+0.09$ for each country's two least divisive dimensions, but equally at every level of attention: the median changes by $+0.001$ from the least to the most interested tercile, and rises at all in 51\% of countries. The same test gives $0.000$ (51\%) for the two least divisive dimensions and $+0.027$ (63\%) for social liberalism with economic left--right, and on the 42 countries where all three pairs can be computed the rises are $+0.025$ for the sharpest pair, $-0.004$ for the least divisive and $+0.035$ for social with economic. Social liberalism with democratic commitment is the sharpest pair in 27 of the 71 countries, so the test rests mostly on that pair. Code: \texttt{27\_attention\_gradient\_top2.py}.

Both of Converse's observations therefore reappear, weak constraint across the ninety-six countries where it can be measured and a rise with attention across the eighty-nine where interest is measured, but the rise is confined to the social--economic pair in the countries where that pair coheres and is absent along the divisions that matter most elsewhere. Constraint that grows with attention is a property of the left--right bundle where it exists rather than a general feature of how citizens organize their views. Where it exists, it is strong and concentrated among the attentive exactly as he described. What does not survive is generalizing from the places where it was established: the countries that exhibit constraint most strongly are mostly wealthy, long-established and programmatically contested, though ten of the 24 above $+0.1$ are not, Pakistan among them at $+0.29$, and the American literature is not wrong about America so much as describing an unusual case.

\subsection*{A benchmark for claims of incoherence}

These estimates also supply a benchmark. Claims that some actor's ideology ``does not hang together'' are made routinely --- about survey respondents, political elites and language models --- and almost always against an unstated standard of what coherence ought to look like. These estimates supply the standard, and it is a distribution rather than a single number: the median country's social--economic correlation is $+0.01$, while the American figure is $+0.54$. An actor whose dimensions correlate at zero is therefore not incoherent by world standards but typical of them, resembling a Brazilian or a Kenyan respondent far more closely than an American one. Which of those comparisons is the appropriate one is a question that the choice of benchmark answers implicitly, and \citet{kaufman2026models} use these values to make that choice explicit.

\FloatBarrier
\section*{Appendix G. Further discussion and limitations}

\subsection*{What individual-level estimates add}

Comparative opinion research has been conducted largely in country means. Estimating people rather than aggregates changes four answers at once. Most political disagreement lies inside countries rather than between them, by a margin --- 58 to 91\% of the variance --- that makes the between-country comparison the smaller part of the problem. Two countries with the same average can differ threefold in how divided they are; the ordering the field reports and the disagreement it seeks to explain are weakly related at best, a country's mean and its spread correlating between $-0.23$ and $+0.24$ across the five dimensions. The demographic categories the field studies sort opinion weakly and inconsistently. Ideological constraint, treated as a feature of mass publics since it was first measured in the United States, proves to be a feature of 24 of the ninety-six countries we can test.

None of those four results can be computed from country means. A share of variance lying within countries cannot be formed from country-level quantities at all. A distribution three times wider than another with the same center is invisible to any statistic that returns a center. A subgroup gap is a difference between people inside a country, and constraint is a relationship between two positions held by the same person; neither exists at the level of an aggregate at all. The methodological claim of this paper is not that individual estimates are more precise than country ones but that they answer questions country estimates cannot be asked. \citet{kaufman2026models} make the same point in a form that is easy to check. Saying that the median model is more socially liberal than 95\% of the world's people, or than three quarters of the population in 138 of 149 countries, requires knowing where the people are and not where the countries are. No amount of additional national auditing would produce those numbers, because the quantity is not identified in that design.

Two of our results are deflationary. Demographic divisions are small, and ideology coheres in far fewer places than the literature assumes. Both run against a large literature that treats education gaps or left--right constraint as settled features of mass opinion, and both are more securely estimated here than in the single-country studies that established them, because they rest on up to 155 countries rather than one. A finding that a widely assumed regularity is weak or absent almost everywhere is not a null result about our measure but a result about the regularity.

The constraint result deserves a stronger statement than ``deflationary''. Our scale detects constraint, and detects it most strongly in the United States, Canada, the Netherlands, Pakistan, Switzerland, and Australia; within the countries where it holds, it rises with political interest exactly as \citet{converse2006nature} predicted. Both halves of his argument replicate. The difficulty was never the theory but that the countries in which it can be observed are also, for unrelated reasons, the countries that are surveyed; a regularity holding in a handful of places has therefore been read as a regularity of mass publics. Distinguishing those two readings requires a measure that reaches the places where the regularity fails, and until now no such measure existed.

\subsection*{Measurement invariance is the central assumption}

The design assumes item parameters are common across countries, so that a given answer means the same thing everywhere. This is standard in cross-national joint scaling, it is the paper's most consequential assumption, and it is not exactly true \citep{king2004enhancing,davidov2014measurement,stegmueller2011apples}. Formal invariance tests on the constructs underlying two of our five dimensions have repeatedly rejected full scalar invariance. The World Values Survey emancipative-values battery is misspecified as a comparable index \citep{sokolov2018index}, its scales of support for democracy can be compared only in part and not across all countries \citep{ariely2011rate}, and its value orientations have been judged comparable only among a small set of advanced post-industrial democracies \citep{aleman2016value}. The one prior item response study spanning the barometer families we use reached the same verdict for political trust \citep{vandermeer2019putting}. Our nearest neighbor in scale carries the term we omit: \citet{claassen2019estimating} adds item-by-country intercepts precisely because item functioning varies with national context.

We therefore screen rather than assert, refitting each dimension within five regional blocks and correlating the resulting item discriminations (Methods). The screen is a coarse diagnostic rather than an estimate of invariance, and agreement varies by dimension and, within dimension, by region. Social liberalism, the dimension carrying most of our headline results, generalizes best, at a median cross-region correlation of 0.79 over the 25 items common to all five blocks; anti-elite sentiment follows at 0.63. Democratic commitment (0.42, on 13 common items) and economic left--right (0.50) are weak, and in both the weakness is African: on democratic commitment Africa's discriminations correlate with the other regions' at $-0.07$ to $0.18$ while the other four agree with each other at 0.27 to 0.73, and on economic left--right the African refit bears no relationship to the pooled fit ($+0.02$), and agrees with the others only at $0.10$ to $0.51$, while the other four agree at 0.46 to 0.85. Country-level quantities on those two dimensions should therefore be read as orderings rather than measurements wherever African publics enter the comparison, and that applies to the countries at the ends of the ordering as much as to those in the middle. Defensible cardinal comparisons come from social liberalism, where the screen is cleanest.

The assumption bears unequally on our results, and the polarization results carry the most weight. As noted in Appendix~E, a comparison of spreads depends on item discriminations. The discrimination screen is strongest on social liberalism, the dimension Figure~\ref{fig:polarization} uses for illustration, and the claim that spread, bimodality, and alignment are three separable quantities is reproduced on each of the other four dimensions. The exposure lies elsewhere: which dimension each country is most divided on compares its spread on social liberalism with its spreads on dimensions whose discriminations travel less well, democratic commitment and economic left--right above all, and the comparison is least secure for African countries, where those two dimensions fail the regional screen. The same caution applies to the Iraq example in the Introduction, which rests on its democratic-commitment percentile. The subgroup results are the least exposed. A gap between a country's graduates and its school-leavers requires an item to function similarly for two groups facing the same questionnaire, not identically across 155 countries, and the same is true of the coherence correlations, which are computed within country throughout.

A second diagnostic bears on the same assumption. Non-invariance across countries is hard to observe directly, because a country appears once and there is nothing to compare it against. Non-invariance across \emph{instruments} is observable, because most of our countries appear in more than one survey program and the programs can be asked whether they agree, and they do not. A survey wave places the same national population up to 1.04 standard deviations from where a wave of another program places it, while consecutive waves of a program repeating its battery differ on average by 0.02 to 0.13. Part of the harmonization therefore does not hold across instruments. The alignment claim, that two items measure the same construct, is approximately rather than exactly true, and the residual appears as an instrument offset. That offset is a difference in location, and the spread model absorbs it by giving each wave its own mean. It runs to 7.6\%, 5.4\%, 4.7\% and 5.3\% of the median country's within-country variance on social liberalism, economic left--right, anti-elite sentiment, and democratic commitment. The two diagnostics therefore measure different failures and do not rank the dimensions alike: social liberalism carries the largest location offsets between instruments but the most stable discriminations across regions, and democratic commitment the reverse. For a spread, which depends on discriminations and not on location once each wave has its own mean, the second is the one that matters. We report the offsets rather than only correcting for them, because an offset of this size produces no error and no diagnostic complaint, appearing instead as one country's public looking unusually divided.

A third diagnostic asks the comparability question item by item, which is closer to the worry that the same answer about democracy, religion, or trust in government means different things in France, Egypt, and China. Within each regional block we correlate every item's responses with the respondents' estimated positions, and call the item failing there if the correlation is below 0.10 or opposite in sign to its pooled value (Methods). Democratic-commitment items fail most, in 18\% of the items tested in Europe, 15\% in the Middle East and North Africa and 14\% in Africa, against 10\% in the Americas and 5\% in Asia; economic left--right items fail at 8 to 16\% in every region. Dropping each failing item only in the blocks where it fails and re-estimating the spreads by maximum likelihood on the fixed parameters leaves country spread rankings correlated with ours at 0.997 to 1.000 and the most divisive dimension unchanged in 143 of the 154 countries (141 of the 146 measured on the same dimensions before and after); dropping the failing items everywhere gives 0.995 to 1.000 and 122 of 154 (102 of 110), the larger change because economic left--right loses 22\% of its answers and with them its spread in 44 countries. The comparison survives removing the items that visibly do not travel. That does not certify that the remaining items mean the same thing everywhere, only that our conclusions do not rest on the ones that visibly fail. Religion in government passes the same screen: no item fails in any region where it can be tested, and its spreads are unchanged. Its items are too specific to their programs for the regional refit.

\subsection*{Where the linkage is thinnest}

Joint scaling works through the items different surveys share, and ours are unevenly distributed. 152 of 1,378 items do all the linking, and the national election studies, our largest block, join the cross-national programs through 40 items, of which 32 carry one of the five dimensions and five the economic one; the Pew survey and the European Election Studies join through two and three items. The corpus is connected, but a scale whose largest block joins the rest through five economic items cannot support a fine-grained economic conclusion. Economic left--right reaches a marginal reliability of 0.74 and an agreement of 0.90 between our two country estimators, close to the other dimensions, but it covers fewer countries than any dimension except religion in government, 110, because the Arab Barometer asks no economic attitude item and the Afrobarometer only one, on the government's responsibility for health care. Continuing that work in the single-country election studies is the single most valuable extension to this corpus.

\subsection*{Observed division is compressed where dissent is dangerous}

The political meaning of a disagreement differs between a society in which people can organize freely around competing positions and one in which opposition is repressed. Polarization in the pernicious sense, a process in which elites and institutions organize disagreement into mutually distrustful camps and which can erode democracy itself \citep{mccoy2018polarization,mccoy2019toward,svolik2019polarization,graham2020democracy}, is shaped by that difference, and attitudinal dispersion is only one input to it. Observed dispersion may also understate division where dissent is dangerous \citep{kuran1995private,tannenberg2022autocratic}, so the lower spread of autocracies could reflect constraint on expression rather than consensus, although sensitivity bias in surveys is usually modest \citep{blair2020worry}. Three checks bear on this (Methods). Spread on social liberalism rises with V-Dem's liberal democracy index (country correlation 0.48; mean spread 0.56 in closed autocracies against 0.85 in liberal democracies), but spread on democratic commitment does not, and spread on anti-elite sentiment falls, although those are the attitudes it is most dangerous to hold in an autocracy, and they should compress most if repression compressed expressed opinion generally. Where constraint does show, it shows on regime-relevant items. Don't-know and refusal run at 11.0\% on democratic-commitment items in closed autocracies against 3.6\% in liberal democracies, rising 1.9 percentage points per standard deviation of lower liberal democracy net of program ($p < 0.001$), with no such gradient on social liberalism; and in autocracies, respondents who say people must be careful about what they say on politics answer democracy items more uniformly than other respondents (median ratio of standard deviations 0.87, lower in 72\% of 24 countries), against 0.96 in democracies. We therefore read spreads on democratic commitment in autocracies as lower bounds. The social-liberalism gap is not a nonresponse artifact, but whether it reflects homogeneous opinion or conformity to an enforced norm is a question survey responses cannot settle.

\subsection*{Other limits}

We report point estimates with standard errors and do not propagate measurement uncertainty through to the downstream comparisons, which in the democratic-support literature has overturned published findings \citep{tai2024democracy}. The limitation bears hardest on the polarization results, where the quantity of interest is itself a variance. Correcting the instrument offsets at source, by estimating item-by-program effects inside the measurement model, would also adjust the country means that the centering described in Methods leaves alone. That model cannot be estimated here. The corpus is close to block-diagonal by construction, which is why it needed a joint scale in the first place, and on social liberalism most items appear in only one program; restricting the model to the items that two or more programs carry leaves it without the anchor items a formal test would need (Methods). The constraint is in the design of the world's survey programs rather than in our estimation.

That is close to a tautology once seen, and it is the limit of this design. A corpus is assembled jointly precisely because its instruments share too few items to be compared directly, and for the same reason it cannot supply the common items a formal test of cross-instrument invariance would need. The assumption that makes the scale possible cannot be tested from inside the scale. We therefore probe it from outside, with the regional screen on discriminations, the decomposition of within-country variance into within- and between-program parts, and the two-way country-by-program model that bounds how far a country's position could move. None of the three is a substitute for an invariance test. Any corpus assembled this way will be in the same position, and its invariance evidence will have to come from external diagnostics rather than from a clean internal test.

Our estimates are cross-sectional, assembled from waves fielded at different times, and nothing here speaks to change over time, the question most of the polarization literature addresses. That limit belongs to this corpus rather than to the method, and it is the one we expect to fall first. Because we estimate the item parameters once and then hold them fixed, a respondent interviewed in 2030 can be scored onto the scale we estimate today; a new wave has only to be harmonized to the existing bank, and its estimates arrive already comparable to the old ones. The programs supplying this corpus field on three- to five-year cycles and are, for the most part, continuing. A second pass at the end of the decade would therefore turn every polarization quantity reported here from a cross-sectional comparison into a trend, measured on one metric across more than a hundred countries. That is why we release the item bank and the fitted parameters rather than the estimates alone.

\subsection*{Issue polarization and the timing of the corpus}

We measure issue polarization, not affective polarization. Nothing in this corpus speaks to how people feel about their opponents, which is the quantity much of the recent literature is about, including its comparative work \citep{gidron2020american,gidron2023who}. Nor do we measure polarization in the pernicious sense, a political process in which elites and institutions organize disagreement into mutually distrustful camps and which can erode democracy itself \citep{mccoy2018polarization,mccoy2019toward,svolik2019polarization,graham2020democracy}. Attitudinal dispersion is one input to that process, and the main text shows how loosely it is tied to the parties that would organize it.

Our estimates are cross-sectional, but the question of widest interest is whether polarization is rising, and that needs a time series this corpus does not have; sources are single waves or a small number of waves, fielded at different times. The 73 survey waves we analyze were fielded between 2013 and 2025, with a median year of 2021 and 57\% of them from 2021 onward; the corpus is thus best read as a snapshot of the late 2010s and early 2020s rather than as a single moment. Six sources carry a fieldwork year we assigned rather than read off a release --- three CSES modules and the Austrian panel, dated to the midpoint of their field period, the British panel, dated to its last wave, and a Danish study whose release states no year --- which affects any analysis grouping by country-period but not the cross-sectional quantities reported here. Where a program contributes multiple waves a within-program trend is estimable, though it describes a few countries rather than the world.

\subsection*{Attitudes and the cleavages that organize conflict}

What this paper measures well is the distribution of particular political attitudes and the relationships among them. Which cleavages structure political conflict in a given society is a broader question, and our results show how unevenly the two meet. Social liberalism is the one dimension on which wide disagreement, issue salience and party sorting go together. Anti-elite sentiment is the dimension parties sort on most, largely because it is the line between government and opposition. And the sectarian divides of Iraq and Lebanon organize party choice almost perfectly while leaving almost no trace in attitudes. Our dimensions are, moreover, the ones that cross-national surveys happen to carry. Religion in government entered only because six cross-national programs and a few election studies ask about it, and ethnic, communal, and regime-identity cleavages enter, where they enter at all, through what people say about corruption, trust, and democracy. When an attitudinal difference becomes a political division --- which disagreements parties choose to organize, and which divisions persist with no disagreement over policy --- is a question these estimates can frame for 155 countries and cannot answer for any one of them. We think it is the most natural next use of a common scale.

Comparative opinion research has spent thirty years comparing countries because that is what the measures allowed, not because countries were the object of interest. The questions the field cares about --- whether publics are pulling apart, who is sorting with whom, whose views are changing and whose are not --- are all questions about the people inside a country, and none of them survive being averaged. The limitation, we have argued, was never in the data. The surveys have been asking individuals all along; they simply asked them different questions in different places, and the arithmetic of putting those answers on one scale had not been done. Once it is, the between-country comparison that has organized the field proves to be the smaller share of the variation, and several things assumed to hold inside countries turn out to hold inside a dozen of them.

\FloatBarrier

\section*{Appendix H. Party sorting, issue salience and party polarization}

The main text compares publics with party systems through party-system polarization. This appendix reports two further ways of asking whether the divisions we measure are the ones politics is organized around: mass-level party sorting, the share of a dimension's within-country variance explained by the party a respondent supports, and issue salience. It closes with the robustness of the party-polarization comparison and with party alignment as an alternative measure of what divides a country.

The first is party. For each country and dimension we compute the share of within-country variance in positions explained by the party a respondent votes for or feels close to, with a correction for the number of parties (Methods). Parties sort anti-elite sentiment most, at a median of 8.0\% of the variance across 143 countries, against 1.5 to 3.5\% on the other four dimensions, and on the raw share anti-elite sentiment is the most party-aligned dimension in 89 of 144 countries. Much of that is the line between government and opposition. In Latinobar\'{o}metro, supporters of the governing party and of the opposition differ on anti-elite sentiment (7.7\% of the variance) and on nothing else (0.3\% or less). The pro- and anti-regime cleavage that students of Middle Eastern and Latin American politics look for is in these data, under that name.

Division and party sorting are related within a dimension and barely across them. Countries more divided on a dimension sort their parties more on it, on social liberalism (Spearman's $\rho = 0.56$) and anti-elite sentiment ($0.39$), weakly on democratic commitment ($0.15$), and the reverse on economic left--right ($-0.38$). The dimension a country is most divided on and the one its parties sort on most coincide in 37\% of the 139 countries where both can be computed, against 23\% by chance, and 48 of the 52 matches are on anti-elite sentiment or social liberalism (last paragraph of this appendix). Religion in government is the most party-aligned dimension in 15 of the 58 countries where it can be compared: Egypt, India, Indonesia, Israel, Jordan, Lebanon, Libya, Malaysia, Morocco, Palestine, Poland, Sri Lanka, Thailand, Tunisia, and Yemen. In the Arab Barometer countries a fuller Islamist--secular battery sorts party more than any other attitude that program carries, at a median of 7.5\% of the variance and 14 to 17\% in Jordan, Lebanon, Tunisia, and Palestine.

Some divisions barely register in attitudes. In Iraq and Lebanon the Sunni--Shia divide explains under 1\% of the variance on democratic commitment and anti-elite sentiment, the two dimensions the Arab Barometer measures in both countries, apart from anti-elite sentiment in Lebanon at 5.1\%, while predicting party almost perfectly (Cram\'{e}r's $V$ of 0.97 in Lebanon and 0.67 in Iraq; Methods). These are publics divided along communal lines with almost no difference in policy views on the dimensions measured, and dispersion measures, ours included, cannot see that kind of division.

The second check is salience. Most programs ask respondents to name the most important problem facing their country, and we code the answers into classes matching our dimensions, using the first mention and any of up to three mentions, for 131 countries (Methods). The answers are dominated by the economy, at a median of 43\% of first mentions, and by public services, at 24\%; these record problems the country faces more than positions on them \citep{wlezien2005salience}, and the most salient dimension matches the most divided one in only 23 to 28\% of countries. Within a dimension, salience, division, and party sorting move together on exactly one. Where more people name social and cultural problems, countries are more divided on social liberalism ($\rho = 0.39$ across 126 countries, on any mention) and their parties sort more on it ($0.59$). Economic salience goes weakly with economic spread on first mentions ($0.27$) but against party sorting ($-0.41$), and democratic and corruption salience are unrelated or negatively related to spread on the corresponding dimensions, with no positive correlation above 0.20 on either measure. Social liberalism is therefore the one dimension on which wide disagreement, issue importance and party sorting line up. On the others, a wide spread shows that people disagree, not that the disagreement is politically organized.

The dimensions themselves are the ones cross-national surveys happen to carry, a narrower set than the cleavages that organize politics outside Western democracies (Appendix~G). Two of those other cleavages are nonetheless visible: the pro- and anti-regime divide appears as anti-elite sentiment, and the Islamist--secular divide as religion in government and in the Arab Barometer battery. Communal divisions that carry no difference in policy views are not visible, and no attitudinal instrument would find them.

\emph{Party polarization: robustness.} V-Party offers two indicators for four of the five dimensions: LGBT social equality and immigration for social liberalism, economic left--right and welfare for economic left--right, the anti-pluralism index and political pluralism for democratic commitment, and anti-elitism and the populism index for anti-elite sentiment; religion in government has religious principles alone. Across party-elections the paired indicators correlate at $0.50$, $-0.47$, $-0.90$ and $0.95$. Used singly, the two sets agree on the dimension a country's parties are most polarized on in only 48\% of the 121 countries, with the disagreement spread across all four dimensions they both measure, where the paired indicators measure related but different things. We therefore average them, oriented by the sign of their correlation, into one position per party. Averaging the two sets' country scores instead, which needs no orientation, agrees with this composite in 87\% of countries. The public's and the parties' sharpest dimension coincide in 45\% of countries under the composite and 42\% under the score average, and in 31\% and 39\% under the two single sets, against 21 to 23\% by chance. Very small party standard deviations, where a system's parties take nearly identical positions, are floored at 5\% of the dimension's median before the log; flooring at 1\% or 10\% instead gives 46\% and 44\%. The near-miss pattern of the third test is weaker under the single indicator sets: the parties' sharpest dimension is among the public's two sharpest in 66\% of countries under the composite and 66\% under the score average, 64\% under the alternative set and 56\% under the main set, and it is the public's least divisive in 17\%, 17\%, 19\% and 21\%, against 24\% by chance. Code: \texttt{22\_party\_polarization\_vparty.R}. The rank correlation between public spread and party polarization on anti-elite sentiment is $-0.39$ under the composite and $-0.40$ and $-0.38$ under the single sets; on social liberalism it is $+0.34$, $+0.19$ and $+0.34$. V-Party ends in 2019, so party positions come from elections between 2010 and 2019, at a median of 2017, against a median respondent fieldwork year of 2019. Thirty-three of the 154 countries lack a usable election in that window, either because V-Party does not cover them or because their most recent elections carry no vote shares; they include China, Vietnam, Jordan, and Lebanon.

\emph{Party alignment as an alternative measure of division.} Spread asks how far a public's members disagree; alignment asks how much of that disagreement runs along party lines, measured as the share of a dimension's within-country variance explained by the party a respondent supports, as above. Applying the main text's rule to alignment instead of spread (the two-way model applied to log alignment, among at least two dimensions) for the 139 countries where it can be computed, anti-elite sentiment is the most party-aligned dimension in 46, social liberalism in 43, democratic commitment in 19, economic left--right in 17, and religion in government in 14. The most party-aligned dimension is the most divided one in 37\% of countries, against 23\% by chance. Part of the regional pattern survives the change of measure. Sub-Saharan Africa is most aligned on anti-elite sentiment in 80\% of its 40 countries, as it is most divided on it; Europe and Central Asia on social liberalism, in 60\% of 45 countries; and Canada and the United States on social liberalism. Elsewhere the two measures part: Latin America is most party-aligned on social liberalism (38\%) and anti-elite sentiment (33\%) rather than economic left--right (14\%), and the Middle East and North Africa on religion in government (64\% of 11 countries). Alignment is also the more portable measure. Where two programs survey the same country, its alignment profile across dimensions correlates at 0.30 (95\% interval 0.20 to 0.40, 73 countries), close to the 0.37 between two waves of the same program, whereas its spread profile correlates at 0.17 against 0.49 (Appendix~E). Code: \texttt{24\_profile\_replication.py}.

\FloatBarrier

\end{document}